%% file: main.tex
\documentclass[conference]{format/IEEEtran_IREP}
\ifCLASSINFOpdf

\else

\fi

\usepackage{cite}
\usepackage{mathtools}
\usepackage{adjustbox}
\usepackage{amssymb,amsfonts,amsmath}
\usepackage{amsthm}
\usepackage{algorithm}
\usepackage{algorithmic}
\usepackage{graphicx,multirow}
\usepackage{textcomp}
\usepackage{color,soul}
\usepackage{enumerate}
\usepackage{pgfplots}
\usepackage{physics}
\usepackage{url}
\usepackage{booktabs,colortbl}
\usepackage{array}
\usepackage{optidef}
\usepackage{bm}
\usepackage{dcolumn}
\usepackage{longtable}
\usepackage{makecell}
\usepackage{adjustbox}
\usepackage[caption=false, font=footnotesize]{subfig}
\usepackage{hyperref}
\usepackage{dblfloatfix}    
\usepackage{setspace}
\usepackage[]{fancyhdr} %
\newcommand{\changefont}{\fontsize{7.5}{7.5}\selectfont}
\IEEEoverridecommandlockouts

\renewcommand{\baselinestretch}{0.985}

\begin{document}

\title{Chance-Constrained AC Optimal Power Flow for Unbalanced Distribution Grids}

\author{\IEEEauthorblockN{Kshitij Girigoudar*, Ashley M. Hou*, and Line A. Roald}
\thanks{*The first two authors contributed equally to this work. K.~Girigoudar, A.~M.~Hou, and L.~A.~Roald are with the Department of Electrical and Computer Engineering, University of Wisconsin-Madison, USA (e-mail: girigoudar@wisc.edu, amhou@wisc.edu,
roald@wisc.edu).}}

\maketitle
\thispagestyle{fancy}
\pagestyle{fancy}

\begin{abstract}

The growing penetration of distributed energy resources (DERs) is leading to continually changing operating conditions, which need to be managed efficiently by distribution grid operators. The intermittent nature of DERs such as solar photovoltaic (PV) systems as well as load forecasting errors not only increase uncertainty in the grid, but also pose significant power quality challenges such as voltage unbalance and voltage magnitude violations. This paper leverages a chance-constrained optimization approach to reduce the impact of uncertainty on distribution grid operation. We first present the chance-constrained optimal power flow (CC-OPF) problem for distribution grids and discuss a reformulation based on constraint tightening that does not require any approximations or relaxations of the three-phase AC power flow equations. We then propose two iterative solution algorithms capable of efficiently solving the reformulation. In the case studies, the performance of both algorithms is analyzed by running simulations on the IEEE 13-bus test feeder using real PV and load measurement data. The simulation results indicate that both methods are able to enforce the chance constraints in in- and out-of-sample evaluations.
\end{abstract}


\input{Sections/0_Nomenclature}
\input{Sections/1_Introduction}
\input{Sections/2_Modelling}

\input{Sections/3_Solution_methods}
\input{Sections/4_Case_study}

\input{Sections/5_Conclusion}

\bibliography{ref}

\end{document}

%% file: Sections/0_Nomenclature.tex
\section*{Nomenclature}

\emph{Sets and Indices}
\begin{IEEEdescription}[\IEEEusemathlabelsep\IEEEsetlabelwidth{$p_{\text{G},i, \omega}^\phi,P_{\text{G},i, \omega}$}]
\item[$\mathcal{N}$] Set of nodes excluding slack bus, $|\mathcal{N}| = n$
\item[$0$] Substation node (slack bus) index
\item[$i,k,l \in \mathcal{N}$] Node index
\item[$\Phi$] Set of phases, $\{a,b,c\}$
\item[$\phi\in \Phi$] Phase index
\item[$\Omega$] Uncertainty set, $|\Omega| = M$
\item[$\omega \in \Omega$] Uncertainty realization 
\item[$\kappa$] Tuning iteration count 
\end{IEEEdescription}
\vspace{6pt}

\emph{Random Variables}
\begin{IEEEdescription}[\IEEEusemathlabelsep\IEEEsetlabelwidth{$p_{\text{G},i, \omega}^\phi,P_{\text{G},i, \omega}$}]
\item[$p_{\text{L},i,\omega}^\phi,P_{\text{L},i,\omega}$] Uncertain active power load demand 
\item[$q_{\text{L},i,\omega}^\phi,Q_{\text{L},i,\omega}$] Uncertain reactive power load demand
\item[$\delta p_{\text{L},i,\omega}^\phi,\delta P_{\text{L},i,\omega}$] Deviation of uncertain active power load demand from average value
\item[$p_{\text{G},i, \omega}^\phi,P_{\text{G},i, \omega}$] Uncertain active power generation of solar PV 
\item[$\delta p_{\text{G},i,\omega}^\phi,\delta P_{\text{G},i,\omega}$] Deviation of uncertain active power generation of solar PV from average value
\item[$p_{i, \omega}^\phi,P_{i, \omega}$] Uncertain active power injection 
\item[$q_{i, \omega}^\phi,Q_{i, \omega}$] Uncertain reactive power injection
\item[$Y_{\overline{v},i}^{\phi}$] Constraint violation indicator random variable
\item[$\hat{E}_{\overline{v}, i}^{\phi}$] Constraint empirical violation probability
\item[$\hat{E}_{\overline{v}}^{\max}$] Worst case empirical violation probability for upper voltage constraint
\end{IEEEdescription}
\vspace{6pt}

\emph{Parameters}
\begin{IEEEdescription}[\IEEEusemathlabelsep\IEEEsetlabelwidth{$|s_{\text{G},i}^\phi|,|S_{\text{G},i}|$}]
\item[${\gamma}_{\text{L},i}^\phi,{\Gamma}_{\text{L},i}$] Power ratio of load demand 
\item[$\overline{p}_{\text{L},i}^\phi,\overline{P}_{\text{L},i}$] Average active power load demand 
\item[$\overline{q}_{\text{L},i}^\phi,\overline{Q}_{\text{L},i}$] Average reactive power load demand
\item[$\overline{p}_{\text{G},i}^\phi,\overline{P}_{\text{G},i}$] Average active power generation of solar PV 
\item[$|s_{\text{G},i}^\phi|,|S_{\text{G},i}|$] Apparent power rating of solar PV 
\item[$\overline{p}_{i, \omega}^\phi,\overline{P}_{i, \omega}$] Average active power injection 
\item[$\overline{q}_{i, \omega}^\phi,\overline{Q}_{i, \omega}$] Average reactive power injection
\item[$\epsilon_q$] Violation probability of solar PV inverter reactive power limits, $\epsilon_q \in [0,1]$
\item[$G,B$] Nodal conductance and susceptance matrices of nodal admittance matrix $Y$
\item[$\underline{v},\overline{v}$] Voltage magnitude limits
\item[$\epsilon_v$] Violation probability of voltage magnitude limits, $\epsilon_v \in [0,1]$
\item[$\overline{q}_{\text{G},i}^\phi,\underline{q}_{\text{G},i}^\phi$] Upper and lower PV inverter limits w.r.t. average generation
\item[$\overline{\lambda}_{q,i}^{\phi}, \overline{\Lambda}_{q}$] Upper reactive power constraint tightening
\item[$\underline{\lambda}_{q,i}^{\phi}, \underline{\Lambda}_{q}$] Lower reactive power constraint tightening
\item[$\overline{\lambda}_{v,i}^{\phi}, \overline{\Lambda}_{v}$] Upper voltage constraint tightening
\item[$\underline{\lambda}_{v,i}^{\phi}, \underline{\Lambda}_{v}$] Lower voltage constraint tightening
\item[$\eta$] Tuning convergence tolerance
\item[$s \in \mathbb{R}^{+}$] Tuning parameters
\item[$s_{\min}, s_{\max}$] Upper and lower tuning parameter limits
\item[$\sigma_{v,i}^{\phi}$] Estimated standard deviation of voltage magnitude

\end{IEEEdescription}
\vspace{6pt}

\emph{Optimization Variables}
\begin{IEEEdescription}[\IEEEusemathlabelsep\IEEEsetlabelwidth{$\boldsymbol{p}_{\mathbf{G},0, \omega}^\phi,\boldsymbol{P}_{\mathbf{G},0, \omega}$}]
\item[$\boldsymbol{q}_{\text{G},i}^\phi,\boldsymbol{Q}_{\text{G},i}$] Reactive power generation of solar PV
\item[$\boldsymbol{p}_{\mathbf{G},0, \omega}^\phi,\boldsymbol{P}_{\mathbf{G},0, \omega}$] Active power injection at substation corresponding to uncertain power injections
\item[$\boldsymbol{q}_{\mathbf{G},0, \omega}^\phi,\boldsymbol{Q}_{\mathbf{G},0, \omega}$] Reactive power injection at substation corresponding to uncertain power injections
\item[$\boldsymbol{p}_{\mathbf{G},0}^\phi,\boldsymbol{P}_{\mathbf{G},0}$] Active power injection at substation corresponding to average power injections
\item[$\boldsymbol{q}_{\mathbf{G},0}^\phi,\boldsymbol{Q}_{\mathbf{G},0}$] Reactive power injection at substation corresponding to average power injections
\item[$|\boldsymbol{v}_{i, \omega}^\phi|,|\boldsymbol{V}_{i, \omega}|$] Voltage magnitude corresponding to uncertain power injections
\item[$\boldsymbol{\theta}_{i, \omega}^\phi,\boldsymbol{\Theta}_{i, \omega}$] Voltage angle corresponding to uncertain power injections
\item[$|\boldsymbol{v}_{i}^\phi|,|\boldsymbol{V}_{i}|$] Voltage magnitude corresponding to average power injections
\item[$\boldsymbol{\theta}_{i}^\phi,\boldsymbol{\Theta}_{i}$] Voltage angle corresponding to average power injections
\item[$\boldsymbol{v}_{\mathbf{d}l, \omega}^-,\boldsymbol{v}_{\mathbf{q}l, \omega}^-$] Real and imaginary components of negative sequence voltage phasor $\boldsymbol{v}_{l, \omega}^-$ corresponding to uncertain power injections
\item[$\boldsymbol{v}_{\mathbf{d}l, \omega}^+,\boldsymbol{v}_{\mathbf{q}l, \omega}^+$] Real and imaginary components of positive sequence voltage phasor $\boldsymbol{v}_{l, \omega}^+$ corresponding to uncertain power injections
\item[$\boldsymbol{v}_{\mathbf{d}l}^-,\boldsymbol{v}_{\mathbf{q}l}^-$] Real and imaginary components of negative sequence voltage phasor $\boldsymbol{v}_{l, \omega}^-$ corresponding to average power injections
\item[$\boldsymbol{v}_{\mathbf{d}l}^+,\boldsymbol{v}_{\mathbf{q}l}^+$] Real and imaginary components of positive sequence voltage phasor $\boldsymbol{v}_{l, \omega}^+$ corresponding to average power injections
\end{IEEEdescription}
\vspace{6pt}

\emph{Functions}
\begin{IEEEdescription}[\IEEEusemathlabelsep\IEEEsetlabelwidth{$p_{\text{G},i, \omega}^\phi,P_{\text{G},i, \omega}$}]
  \item[$\mathbb{P}_{\omega}(\cdot)$] Chance constraint probability
  \item[$\text{C}(\cdot),\text{S}(\cdot)$] Cosine and sine components of branch angle matrix
  \item[$f(\cdot)$]  Power flow equations (short-hand)
  \item[$f_{q,i}^{\phi}(\cdot)$]  Empirical distribution of reactive power limit
  \item[$f_{v,i}^{\phi}(\cdot)$]  Empirical distribution of voltage limit
\end{IEEEdescription}
\vspace{6pt}

\vspace{6pt}

%% file: Sections/1_Introduction.tex
\section{Introduction}


The increasing integration of distributed energy resources (DERs), such as rooftop solar photovoltaic (PV) systems and electric vehicles, poses both operational challenges and opportunities for distribution grid operations. Large scale penetration of DERs can lead to increases in power injection uncertainty and load variability. This is particularly challenging at the distribution level, where DERs are typically not equally distributed throughout the feeder.
At the same time, the presence of DERs provide new opportunities for control, including using offline design of individual voltage droop control curves~\cite{eggli2020stability,antoniadou2017distributed}, limited communication of real-time measurements~\cite{turitsyn2010distributed,yao2020mitigating}, or direct dispatching of inverters based on frequent resolving of centralized optimal power flow (OPF) problems~\cite{bajo2015voltage,karagiannopoulos2018centralised,su2014optimal,girigoudar2021linearized}. 
While centralized OPF provides optimal set-points, limited real-time system measurements and communication delays can impact the frequency at which the set-points are updated. For example, smart meters can typically take measurements every 15 minutes, but this data is often only available to system operators with several hours delay~\cite{alimardani2015distribution}. As a result, it may be important to identify control set-points ensure system security for extended periods of time, with intervals ranging from 15 minutes (if good communication systems exist) to an entire season (if low complexity is desired). Significant variability and uncertainty in the load and DER power injections is present across these time horizons, which, if not appropriately accounted for, can lead to constraint violations and high voltage unbalance between control set-point updates.
To mitigate this problem, we propose treating the variable load and DER injections as uncertain and formulating the centralized OPF problem as a stochastic optimization problem.

Approaches for stochastic OPF with uncertain renewable energy generation and load
have been well studied in the context of transmission systems~\cite{zhang2011chance,roald2013analytical,summers2015stochastic,summers2014stochastic,liang2013two}. However, methods developed for transmission systems
often leverage DC power flow representations and single-phase equivalents to represent the system, which are not applicable to distributions grids, who are
inherently unbalanced and exhibit high R/X ratios. Furthermore, while transmission operations focus primarily on congestion management and system balancing, distribution utilities focus on managing voltage magnitudes, voltage unbalance, and other power quality issues for end customers. As a result, there is a range of stochastic OPF models that have been developed specifically for distribution grids. This includes robust and distributionally robust methods~\cite{chen2018robust, soares2017active, mieth2018data}, stochastic approximation techniques~\cite{kekatos2015stochastic}, and chance-constrained formulations~\cite{karagiannopoulos2017operational}. These models focus primarily on managing or reducing voltage magnitude violations~\cite{chen2018robust,mieth2018data} while minimizing objectives such as cost of energy~\cite{soares2017active,karagiannopoulos2017operational,mieth2018data}, deviation from a desired power withdrawal at the substation~\cite{mieth2018data}, or losses~\cite{kekatos2015stochastic,chen2018robust,karagiannopoulos2017operational}. They often leverage power flow formulations that rely on a radial network topology~\cite{sankur2016linearized,bernstein2018load, jabr2006radial,gan2014exact,kersting2006distribution}, include approximate representations of unbalance~\cite{arnold2016optimal,dall2013distributed,lavaei2011zero,zhao2017convex}, and sometimes take advantage of iterative solution algorithms such as forward-backward sweep~\cite{karagiannopoulos2018centralised,fortenbacher2016optimal,karagiannopoulos2019data}. Many of these formulations are linear approximations~\cite{sankur2016linearized,arnold2016optimal,dall2017chance} or convex relaxations~\cite{jabr2006radial,gan2014exact,dall2013distributed,lavaei2011zero,zhao2017convex}, which may converge to solutions that are not actually AC feasible. Methods that do consider the full, non-linear, non-convex AC power flow formulation~\cite{soares2017active} or guarantee convergence to a solution that satisfies those equations~\cite{karagiannopoulos2017operational} typically only consider balanced, single-phase systems.

In this paper, we take a more comprehensive view on voltage management than previous work and focus specifically on minimizing voltage unbalance, which causes damage to three-phase motors~\cite{muljadi1985induction}. 
Existing approximate models~\cite{arnold2016optimal,dall2013distributed,lavaei2011zero,zhao2017convex} are inadequate for a detailed analysis of voltage unbalance, as they assume that the system is nearly balanced to start with. Instead, we build our optimization problem around a full, detailed representation of the three-phase power flow, including single-phase, two-phase, and untransposed three-phase lines that serve unbalanced loads~\cite{kersting2006distribution}. 

To minimize voltage unbalance while ensuring that voltage magnitude constraints and DER capacity limits are satisfied, we adopt a chance-constrained formulation. In this model, we ensure that the voltage magnitude and inverter limits are enforced with high probability using single chance constraints.
Chance constraints have the benefit of offering an intuitive trade-off between optimality (i.e., lowering voltage unbalance) and system robustness or reliability (i.e., limiting constraint violations) by adjusting the desired violation probability parameter.
We note an important distinction between voltage and inverter limits: voltage limits are soft constraints, i.e., small and infrequent voltage violations may be acceptable and there is no inherent mechanism in place to enforce that they are always within bounds, while inverter constraints are hard limits, i.e., inverters have control systems in place to ensure that the apparent power from the inverter always stays at or below the maximum threshold as specified by the IEEE Standard 1547-2018~\cite{photovoltaics2018ieee}. Therefore, voltage constraint violations reflect actual over- or under-voltages in the system, while inverter constraint violations reflect a situation in which the inverters will provide less apparent power than our system model assumes. A better system model therefore incorporates the capping of inverter power, i.e., curtailing the inverter reactive power to stay with the limit. However, modeling this type of capping behavior within the optimization problem can be challenging.

In general, chance-constrained optimization problems are hard to solve~\cite{geng2019data}, as the probabilistic constraints must be reformulated to obtain a tractable representation. Methods to reformulate constraints in distribution grid applications include moment-based reformulations assuming Gaussian uncertainty~\cite{cao2013chance} as well as distributionally agnostic methods via the use of, e.g., conservative convex approximations~\cite{dallanese2016optimal,baker2016distribution}, scenario approaches~\cite{venzke2018convex}, or data-driven robust methods~\cite{mieth2018data}.
In this work, we leverage two data-driven, iterative approaches to solving AC chance-constrained optimal power flow (CC-OPF), inspired by existing methods for transmission networks~\cite{schmidli2016stochastic, roald2017chance, roald2017power, hou2020chance} and balanced distribution networks \cite{karagiannopoulos2017operational}.
The key idea is to represent each chance constraint using the nominal constraint with the addition of a tightening term (also known as an uncertainty margin) that secures the constraint against uncertainties. An optimally chosen tightening term will yield solutions that satisfy the chance constraints with the exact desired probability level. The methods iterate between solving an approximate, deterministic OPF with fixed tightening terms and using available uncertainty data to evaluate the resulting solution and update the tightening terms.
A more principled approach to updating or tuning the tightening terms was developed in~\cite{hou2021data} and applied to transmission networks formulated with a DC linearized power flow approximation.
This tuning-based approach solves a sequence of simpler optimization problems, resulting in better computationally tractability, while leveraging detailed system models in the evaluation process, where we simulate system behavior, to allow for the integration of more complex behavior such as inverter capping.


A particular challenge for stochastic and data-driven methods in distribution grids is the characterization of the uncertainty. Assumptions that may be true when considering large numbers of customers (e.g., the law of large numbers) no longer apply to the small number of consumers in distribution feeders, where both load and DER power injections are hard to accurately forecast and exhibit large variability over time. The most flexible way of modeling these distributions is using scenarios. 
However, the quality of the solution may be sensitive to the choice of scenarios, and therefore choosing an appropriate set of samples (e.g., based on historical data) is a challenging yet important aspect that often is overlooked. To remedy this, we investigate how the number of data points and choice of scenario sets impact the solution quality. Using realistic load and solar PV data from Pecan Street~\cite{street2015dataport}, we assess how the solution changes as we use either entire days of data or randomly sampled data points across multiple days.


In summary, the contributions of the paper are as follows:
\begin{enumerate}
    \item We present a chance-constrained formulation of the three-phase AC OPF problem which minimizes voltage unbalance by controlling inverter set-points. Our formulation extends existing work in the literature by considering the full three-phase, unbalanced AC power flow physics and incorporating the effect of inverter capping.
    
    \item We propose a computationally tractable reformulation of our chance-constrained problem, extending the idea of uncertainty margins from~\cite{schmidli2016stochastic, roald2017chance} to a distribution grid setting. We then compare two data-driven algorithms for identifying appropriate constraint tightening values: directly using the results of a Monte Carlo simulation (inspired by transmission grid AC CC-OPF in~\cite{roald2017chance}) or tuning a safety factor (inspired by a transmission grid DC CC-OPF in~\cite{hou2020chance}). Relative to prior work, these updated algorithms include a number of adaptations to make them applicable to the unbalanced distribution grid setting.
    
    \item We perform a detailed case study using real load and PV data, where we investigate three important aspects of data-driven chance-constrained optimization in distribution grids:
    \begin{enumerate}
        \item We investigate how properties of the uncertainty data (i.e., sampling method and correlation between samples) can influence the resulting solutions.
        \item We compare our two methods with respect to their ability to enforce chance constraint satisfaction.
        \item We assess whether the proposed methods are able to effectively integrate the effect of hard PV inverter limits via the use of inverter capping.
    \end{enumerate}
\end{enumerate}

The remainder of the paper is organized as follows: Section \ref{sec:problem-formulation} presents the uncertainty modeling and formulation of the AC CC-OPF problem. Section \ref{sec:solution-methods} details the analytical reformulation and also provides a description of the iterative quantile-based and tuning-based solution algorithms. We introduce the test case and data sets used in the case study in Section \ref{sec:case-study} and present the numerical results and analysis in Sections \ref{sec:case-study-1},~\ref{sec:single-rep-results} and \ref{sec:case-study-2}. Section \ref{sec:conclusion} concludes. 

%% file: Sections/2_Modelling.tex
\section{Problem Formulation}
\label{sec:problem-formulation}

In this section, we describe our model of the PV inverters and load in the distribution grid. We then present a formulation of the three-phase chance-constrained AC OPF problem. 

\subsection{Notation}
We use the following phasor notation:~$X = |X|\angle{\theta}=X_\text{d}+jX_\text{q}$, where~$X$~is a complex phasor, ~$|X|$~and~$\theta$~represent the magnitude and angle components, ~$X_\text{d}$ and $X_\text{q}$~denote the real and imaginary components, and $j=\sqrt{-1}$. All optimization variables are denoted using \textbf{bold} symbols. We use a subscript $\omega$ to represent dependency on a given uncertainty realization $\omega$ (discussed in detail in Section~\ref{sec:uncertainty}).
All scalar values are denoted using small letters, while all vector counterparts and matrices are denoted using capital letters. We represent the element-wise product of two vectors using~$\odot$.

\subsection{Uncertainty Modeling}
\label{sec:uncertainty}
\begin{figure}[!b]
	\centering	        	
	\includegraphics[width=0.33\textwidth]{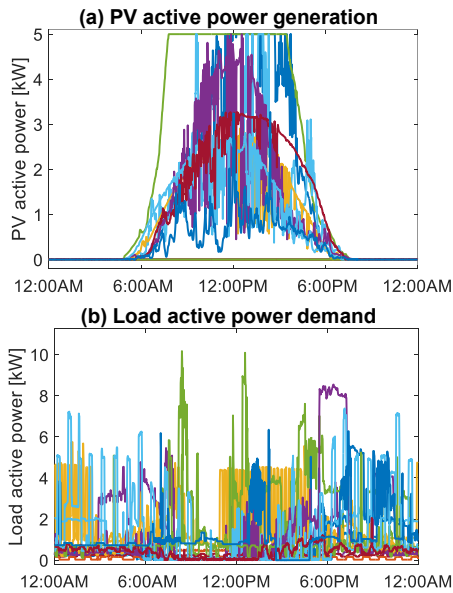}
	\caption{One-minute resolution Pecan Street data for 15 houses from a single day. (a)  PV active power generation. (b) Load active power demand.}
	\label{fig:meas_data}
\end{figure}

Due to the small number of households served by different parts of a distribution feeder, we do not observe the same smoothing effect as in transmission grids~\cite{kersting2006distribution}. The electricity consumption and PV production of individual households is highly variable, as illustrated by the data in Fig.~\ref{fig:meas_data} with measurements obtained from Pecan Street for a set of households in New York~\cite{street2015dataport}. Not only is the load very challenging to forecast, but the data is not necessarily well described by standard probability distributions. To capture the realistic probability distributions and dependence structure of the solar PV and loads, we propose to directly leverage historical data (which may, in many cases, be obtained by utilities from smart meters after a delay~\cite{alimardani2015distribution}).
We denote the full data set as $\Omega$, with individual realizations represented as $\omega \in \Omega$. 
This sample set may include generation and load across certain days (i.e., representative days) or randomly drawn samples from a larger set of historical data (where the resulting sample set has no temporal structure). 

We next describe our load and PV modeling assuming a network where $\mathcal{N}$ denotes the set of three-phase nodes and~$\Phi=\{a,b,c\}$ represents the set of phases.

\subsubsection{Modeling of loads}
For each node $i\in\mathcal{N}$, we denote the three-phase active and reactive power consumption for a given realization $\omega$ as $P_{\text{L},i, \omega}=[p_{\text{L},i, \omega}^{a}~p_{\text{L},i, \omega}^{b}~p_{\text{L},i, \omega}^{c}]^\top$ and $Q_{\text{L},i,\omega}=[q_{\text{L},i, \omega}^a~q_{\text{L},i, \omega}^b~q_{\text{L},i, \omega}^c]^\top$, respectively. 
If reactive power measurements are not available (as will be the case in our case study), we assume that the loads operate with a constant power factor~$pf_i^{\phi}$, giving rise to the following reactive power consumption:
\begin{align}
    & Q_{\text{L},i,\omega} ={\Gamma}_{\text{L},i} \odot {P}_{\text{L},i,\omega}, & \forall i \in \mathcal{N}, \omega \in \Omega,
\end{align}
where the constant factor~${\Gamma}_{\text{L},i} = [{\gamma}_{\text{L},i}^a~ {\gamma}_{\text{L},i}^b~ {\gamma}_{\text{L},i}^c]^\top$ is computed based on the power factor~$pf_i^{\phi}$ with
\begin{align*}
    & {\gamma}_{\text{L},i}^\phi=\sqrt{(1-{pf_i^\phi}^2)/({pf_i^\phi}^2)}, & \forall \phi \in \Phi, i \in \mathcal{N}.
\end{align*}
We define the average active load~$\bar{P}_{\text{L},i}=[\bar{p}_{\text{L},i}^a~\bar{p}_{\text{L},i}^{b}~\bar{p}_{\text{L},i}^{c}]^\top$ and average reactive load~$\bar{Q}_{\text{L},i}=[\bar{q}_{\text{L},i}^a~\bar{q}_{\text{L},i}^{b}~\bar{q}_{\text{L},i}^{c}]^\top$ as the sample average across all realizations in our sample set $\Omega$ with
\begin{subequations}
\begin{align}
    \bar{p}_{\text{L},i}^\phi &= \frac{1}{|\Omega|} \sum_{\omega\in\Omega} p_{\text{L},i, \omega}^\phi, & \forall \phi \in \Phi, i \in \mathcal{N}, \\
    \bar{q}_{\text{L},i}^\phi &= \frac{1}{|\Omega|} \sum_{\omega\in\Omega} q_{\text{L},i, \omega}^\phi,
    & \forall \phi \in \Phi, i \in \mathcal{N},
\end{align}
\end{subequations}
and express the uncertainty of the load consumption as deviations from this average,
\begin{subequations}
\begin{align}
    \delta P_{\text{L},i,\omega} &= P_{\text{L},i, \omega}- \bar{P}_{\text{L},i}, & \forall i \in \mathcal{N},\omega \in \Omega, \\
    \delta Q_{\text{L},i,\omega} &= Q_{\text{L},i, \omega}- \bar{Q}_{\text{L},i} & \forall i \in \mathcal{N},\omega \in \Omega.
\end{align}
\end{subequations}

\subsubsection{Modeling of active power generation from solar PV}
The active power generation of the solar PV inverters under realization $\omega$ is defined as $P_{\text{G},i,w}=[p_{\text{G},i, \omega}^a~p_{\text{G},i, \omega}^b~p_{\text{G},i, \omega}^c]^\top$ and we define the average generation~$\bar{P}_{\text{G},i}=[\bar{p}_{\text{G},i}^a~\bar{p}_{\text{G},i}^{b}~\bar{p}_{\text{G},i}^{c}]^\top$ and uncertain deviation~$\delta P_{\text{G},i,\omega}$ in a similar way as for the loads, i.e.,
\begin{subequations}
\begin{align}
    &\bar{p}_{\text{G},i}^\phi \!=\! \frac{1}{|\Omega|}\!\sum_{\omega\in\Omega} p_{\text{G},i, \omega}^\phi, &\forall \phi \in \Phi, i\!\in\! \mathcal{N}, \\
    &\delta P_{\text{G},i,\omega} \!= \!P_{\text{G},i, \omega}\!-\!\bar{P}_{\text{G},i}, &\forall i\!\in\! \mathcal{N},\omega \in \Omega.
\end{align}
\end{subequations}
We assume that the utility does not wish to curtail active power and therefore consider the active power of the PV inverters as uncontrollable random variables. 

\subsection{Reactive Power Control from Solar PV Inverters}
\label{sec:inverter-control}
We assume that all solar PV inverters are equipped with smart inverters and provide opportunities for reactive power control as outlined in the IEEE Standard 1547-2018~\cite{photovoltaics2018ieee}. For the purposes of this paper, we assume that the inverter is operating in the constant reactive power mode, where it provides reactive power according to a given set-point,
\begin{align*}
    & \boldsymbol{Q}_{\mathbf{G},i}=[\boldsymbol{q}_{\mathbf{G},i}^a~\boldsymbol{q}_{\mathbf{G},i}^b~\boldsymbol{q}_{\mathbf{G},i}^c]^\top, & \forall i\in \mathcal{N}.
\end{align*}
This setpoint is provided by the utility (without accounting for, e.g., local voltage measurements) and serves as a decision variable in our problem. Our goal is to identify a suitable set-point $\boldsymbol{Q}_{\mathbf{G},i}$ that will remain the same for all uncertainty realizations $\omega$ and contributes to minimizing voltage unbalance while keeping voltage magnitudes within bounds.
Note that other types of reactive power control, such as the voltage droop-control outlined in~\cite{photovoltaics2018ieee}, could also be considered but is deferred to future work. 

We further assume that the active and reactive power of the PV inverter must adhere to limits on apparent power. 
For a single-phase PV inverter connected to phase~$\phi$~at node~$i$ with apparent power rating~$|s_{\mathrm{G},i}^\phi|$,  
the inverter reactive power~$\boldsymbol{q}_{\mathbf{G},i}^\phi$ is constrained by
\begin{align}
\label{eq:inv_pow}
    &\mathbb{P}_\omega\! \big( (\boldsymbol{q}_{\mathbf{G},i}^\phi)^2\! +\! (p_{\mathrm{G},i,\omega}^\phi)^2 \!\leq \!|s_{\mathrm{G},i}^\phi| \big) \!\geq \! 1\!-\!\epsilon_q,~\forall_{\phi \in \Phi, i\in \mathcal{N},\omega \in \Omega}.
\end{align}
Because the active power generation $p_{\text{G},i,\omega}^\phi$ varies according to uncertainty realizations $\omega$, this constraint is enforced as a chance constraint with acceptable violation probability $\epsilon_q \in [0,1]$. The chosen reactive power set-point $\boldsymbol{q}_{\mathbf{G},i}^\phi$ is guaranteed to be feasible with probability $1-\epsilon_q$. 

\begin{figure}[!t]
	\centering	        	
	\includegraphics[width=0.28\textwidth]{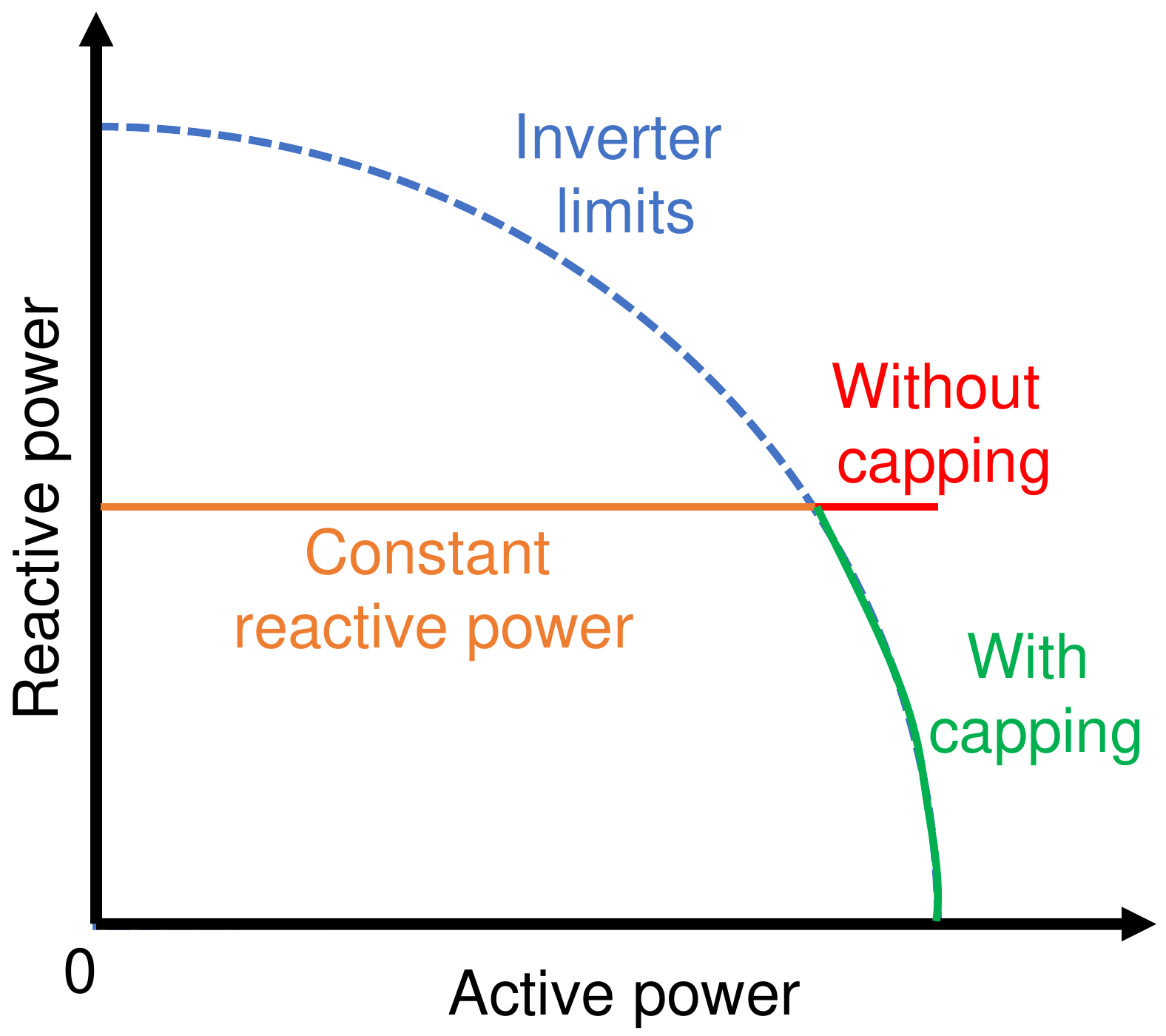}
	\caption{Inverter reactive power control with and without capping.}
	\label{fig:capping}
\end{figure}

Fig.~\ref{fig:capping} illustrates the PV inverter limit (blue dashed line), which is assumed to be a hard constraint, and the reactive power is constant (orange line) with respect to the active power. If the reactive power remains fixed beyond a certain level of active power generation, the PV inverter is overloaded and operates in the region where the inverter limits are violated, indicated by the red line. Operating under overloaded conditions for long duration of time can lead to premature failure of the inverter, potentially requiring manual intervention. To avoid this, we consider capping the inverter to be within the specified limits by reducing the reactive power, as shown by the green curve. This also complies with the IEEE Std 1547-2018~\cite{photovoltaics2018ieee}, which states that DERs are required to respect their apparent power limits. For such cases, the violation probability $\epsilon_q$ should be interpreted as the probability that an inverter is not able to provide the desired reactive power to the grid.

\subsection{Grid Model}
We consider a three-phase distribution grid with one slack node and a set of remaining nodes $\mathcal{N}$, with $n=|\mathcal{N}|$. While we limit our case study to radial distribution grids, our model makes no assumptions of radiality and could thus also be applied to distribution grids that are operated in non-radial configurations. Distribution grids typically have several single and two-phase nodes. However, to simplify notation, we consider all nodes to have three phases. For the single and two-phase nodes, the corresponding entries for the missing phases are set to zero. The resulting total number of single-phase connections in the distribution grid is~$3(n+1)$.

The distribution substation is chosen as the slack node with index~$i=0$. 
We assume that there is one solar PV inverter and one load at each single-phase connection of every node~$i \in \mathcal{N}$. 
If any node connected to a phase has no source or load, we set the corresponding entries to zero.

\subsubsection{Grid parameters}
Following~\cite{girigoudar2020impact}, the critical distribution grid components (including distribution lines, cables, and transformers) are modeled using an overall nodal admittance matrix
\begin{align}
    Y = G + j B \in \mathbb{C}^{3(n+1) \times 3(n+1)}, \label{eq:Ynode}
\end{align}
where $G,~B \in \mathbb{R}^{3(n+1) \times 3(n+1)}$ denote the nodal conductance and susceptance matrices, respectively.

\subsubsection{Nodal power injections}
The active power injection~${{P}}_{i,\omega}\in \mathbb{R}^{3}$ at node~$i \in \mathcal{N} $ varies with the uncertainty realization $\omega$. It is modeled as the sum of the average active generation~$\bar{P}_{\text{G},i} \in \mathbb{R}^{3}$ and average load~$\bar{P}_{\text{L},i}\in \mathbb{R}^{3}$ with corresponding deviations $\delta P_{\text{G},i,\omega} \in \mathbb{R}^{3} \text{ and } \delta P_{\text{L},i,\omega} \in \mathbb{R}^{3}$,
\begin{align}
     {P}_{i,\omega}
     \!=\! \underbrace{\left(\bar{P}_{\text{G},i}+{\delta  P}_{\text{G},i,\omega}\right)}_{{P}_{\text{G},i,\omega}} - \underbrace{\left({\bar{P}_{\text{L},i} +{\delta P}_{\text{L},i,\omega}}\right)}_{{P}_{\text{L},i,\omega}}, ~ \forall_{i \in \mathcal{N},\omega \in \Omega}. \label{eq:P_uc} 
\end{align}
The reactive power injection~${{Q}}_{i,\omega} \in \mathbb{R}^{3}$ at node~$i \in \mathcal{N} $ is modeled as the difference between the controllable reactive power generation of solar PV inverters~${\boldsymbol{Q}}_{\mathbf{G},i}$ and the reactive power load~${{Q}}_{\text{L},i,\omega}$, i.e., ${Q}_{i,\omega} = {\boldsymbol{Q}}_{\mathbf{G},i} - {Q}_{\text{L},i,\omega}$. The average active and reactive power injections at each node are defined as $\bar{P}_{i} = {\bar{P}}_{\text{G},i} - \bar{P}_{\text{L},i}$ and $\bar{Q}_{i} = {\boldsymbol{Q}}_{\mathbf{G},i} - \bar{Q}_{\text{L},i}$, respectively.

The active power balance is maintained by the substation, which supplies the difference between the active power generation and load for each scenario as well as any additional power needed 
to cover the power losses. 
We treat the active power injection at the substation as a decision variable ${\boldsymbol{P}}_{\mathbf{G},0,\omega}$.
Similarly, the substation node also guarantees reactive power balance. 
We define the reactive power injection at the substation as a decision variable $\boldsymbol{Q}_{\mathbf{G},0,\omega}$, which covers the difference between production, consumption, and losses.

\subsubsection{Voltage representation}
The three-phase OPF is implemented in the polar coordinate frame using the phase-to-neutral voltage magnitude and angle variables at every node~$i\in\mathcal{N}$. We denote the voltage magnitudes and angles corresponding to the average power injections $\bar{P}_i,~\bar{Q}_i$ using~$|\boldsymbol{V}_i|=[|\boldsymbol{v}_i^a| ~|\boldsymbol{v}_i^b| ~|\boldsymbol{v}_i^c|]^\top$ and $\boldsymbol{\Theta}_i=[\boldsymbol{\theta}_i^a ~\boldsymbol{\theta}_i^b ~\boldsymbol{\theta}_i^c]^\top$, respectively. As the active and reactive power injections change, the voltages change as well. The voltage magnitude and angle for a given realization $\omega$ is denoted by 
$|\boldsymbol{V}_{i,\omega}|=[|\boldsymbol{v}_{i,\omega}^a| ~|\boldsymbol{v}_{i,\omega}^b| ~|\boldsymbol{v}_{i,\omega}^c|]^\top$ and $\boldsymbol{\Theta}_{i,\omega}=[\boldsymbol{\theta}_{i,\omega}^a ~\boldsymbol{\theta}_{i,\omega}^b ~\boldsymbol{\theta}_{i,\omega}^c]^\top$.
The distribution substation is considered as the reference for voltage angle measurements. We assume the voltage is independent of~$\omega$ and fixed at the substation, i.e.,
\begin{align}
        |{\boldsymbol{V}}_{0}|\angle {\boldsymbol{\Theta}}_{0} = 
    {\begin{bmatrix}
        {1\angle 0^\circ} & {1\angle-120^\circ} & {1\angle 120^\circ}
    \end{bmatrix}}^\top.  \label{eq:Vref}
\end{align}
All other voltage magnitudes are constrained by
\begin{align}
    &\mathbb{P}_\omega \big( |{\boldsymbol{v}}_{i,\omega}^\phi| \leq \overline{v} \big) \geq 1 - \epsilon_{v},  & \forall \phi \in \Phi, i \in \mathcal{N},\omega \in \Omega, \\
    &\mathbb{P}_\omega \big( |{\boldsymbol{v}}_{i,\omega}^\phi| \geq \underline{v} \big) \geq 1 - \epsilon_{v}, & \forall \phi \in \Phi, i \in \mathcal{N},\omega \in \Omega,  \label{eq:Vlim}
\end{align}
where~$\underline{v}$ and~$\overline{v} \in \mathbf{R}$ are the respective lower and upper voltage magnitude bounds. Because the voltage magnitude $|\boldsymbol{v}_{i,\omega}^\phi|$ are dependent on the uncertainty realization $\omega$, these constraints are enforced as with acceptable violation probability $\epsilon_v \in [0,1]$. 
The voltage magnitude constraints can be considered as \emph{soft} constraints, where a constraint violation indicates an under- or over-voltage. The violations of soft constraints is acceptable if the duration and magnitude are small.

\subsubsection{Power flow}
Following~\cite{girigoudar2020impact}, we express the power balance equation at node $i \in \{0,\mathcal{N}\}$ using
\begin{subequations}
\label{eq:pow_bal}
\begin{align}
     {P}_{i,\omega} = 
     &|{\boldsymbol{V}}_{i,\omega}| \odot\sum_{k \in \{0,\mathcal{N}\}}\Big[G_{ik}\odot \text{C}\big(\!{\boldsymbol{\Theta}}_{ik,\omega}\big)\Big] |{\boldsymbol{V}}_{k,\omega}|  \notag\\
     + &|{\boldsymbol{V}}_{i,\omega}| \odot\sum_{k \in \{0,\mathcal{N}\}}\Big[B_{ik}\odot \text{S}\big({\boldsymbol{\Theta}}_{ik,\omega}\!\big) \Big] |{\boldsymbol{V}}_{k,\omega}|,  \\
    {Q}_{i,\omega} =
     &|{\boldsymbol{V}}_{i,\omega}|\odot\sum_{k \in \{0,\mathcal{N}\}}\Big[G_{ik}\odot \text{S}\big({\boldsymbol{\Theta}}_{ik,\omega}\big) \Big] |{\boldsymbol{V}}_{k,\omega}|  \notag \\
     - &|{\boldsymbol{V}}_{i,\omega}|\odot\sum_{k \in \{0,\mathcal{N}\}}\Big[B_{ik}\odot \text{C}\big({\boldsymbol{\Theta}}_{ik,\omega}\big) \Big] |{\boldsymbol{V}}_{k,\omega}|,  
\end{align} 
\end{subequations}
where~$G_{ik},~B_{ik} \in \mathbb{R}^{3\times 3}$ represent the real and imaginary sub-matrices of the nodal admittance matrix~$Y$ for a three-phase branch~$ik$. We represent the cosine and sine components of the branch angle matrix~$\boldsymbol{\Theta}_{ik,\omega}\in \mathbb{R}^{3\times 3}$ using~$\text{C}({\boldsymbol{\Theta}}_{ik,\omega})$ and~$\text{S}({\boldsymbol{\Theta}}_{ik,\omega})$, respectively.
Note that the equations~\eqref{eq:pow_bal} represents the AC power flow equations for all~$\omega \in \Omega$.
For the remainder of the paper, we will use
\begin{align}
    & f(|\boldsymbol{V}_{i,\omega}|,\boldsymbol{\Theta}_{i,\omega},{{P}}_{i,\omega},{{Q}}_{i,\omega})=0, & \forall i \in \{0,\mathcal{N}\},~\omega\in \Omega, \notag
\end{align}
as a shorthand representation of the power balance constraints defined in~\eqref{eq:pow_bal}.


\subsection{Objective Function} 
Our objective for the OPF problem is to minimize voltage unbalance. We utilize the IEC standard~\mbox{61000-2-2}~\cite{standard2002} commonly referred to as Voltage Unbalance Factor (VUF) to define voltage unbalance. For a three-phase node~$l$, the square of VUF for the uncertainty realization~$\omega$ is expressed as
\begin{align}
     {\text{VUF}}_{l,\omega}^2= \frac{{|{v}_{l,\omega}^-|}^2}{{|{v}_{l,\omega}^+|}^2} =   \frac{{({\boldsymbol{v}}_{\mathbf{d}l,\omega}^-)}^2+{({\boldsymbol{v}}_{\mathbf{q}l,\omega}^-)}^2}{{({\boldsymbol{v}}_{\mathbf{d}l,\omega}^+)}^2+{({\boldsymbol{v}}_{\mathbf{q}l,\omega}^+)}^2}, \label{eq:VUF_obj}
\end{align}
where~${\boldsymbol{v}}_{\mathbf{d}l,\omega}^-,{\boldsymbol{v}}_{\mathbf{q}l,\omega}^-$~and~${\boldsymbol{v}}_{\mathbf{d}l,\omega}^+,{\boldsymbol{v}}_{\mathbf{q}l,\omega}^+$~are the rectangular form representation of negative sequence voltage~${v}_{l,\omega}^-$~and positive sequence voltage~${v}_{l,\omega}^+$, respectively, and expressed as non-linear functions of our voltage variables~$\boldsymbol{V}_{l,\omega},\boldsymbol{\Theta}_{l,\omega}$ using
\begin{subequations}
\label{eq:Vn_Vp}
\begin{align}
    &{\boldsymbol{v}}_{\mathbf{d}l,\omega}^- = \Re\{{v}_{l,\omega}^-\}, {\boldsymbol{v}}_{\mathbf{q}l,\omega}^- = \Im\{{v}_{l,\omega}^-\},~\text{where} \\
     &{v}_{l,\omega}^- \!=\! |{\boldsymbol{v}}_{l,\omega}^a|\angle{{\boldsymbol{\theta}}_{l,\omega}^a} \!+\! |{\boldsymbol{v}}_{l,\omega}^b|\angle({{\boldsymbol{\theta}}_{l,\omega}^b\!-\!120^\circ}) \!+\! |{\boldsymbol{v}}_{l,\omega}^c|\angle({{\boldsymbol{\theta}}_{l,\omega}^c\!+\!120^\circ}),  \notag\\
     &{\boldsymbol{v}}_{\mathbf{d}l,\omega}^+ = \Re({v}_{l,\omega}^+), {\boldsymbol{v}}_{\mathbf{q}l,\omega}^+ = \Im({v}_{l,\omega}^+),~\text{where} \\
     &{v}_{l,\omega}^+ \!=\! |{\boldsymbol{v}}_{l,\omega}^a|\angle{{\boldsymbol{\theta}}_{l,\omega}^a} \!+\! |{\boldsymbol{v}}_{l,\omega}^b|\angle({{\boldsymbol{\theta}}_{l,\omega}^b\!+\!120^\circ}) \!+\! |{\boldsymbol{v}}_{l,\omega}^c|\angle({{\boldsymbol{\theta}}_{l,\omega}^c\!-\!120^\circ}). \notag
\end{align}
\end{subequations}

\subsection{Chance-Constrained Optimal Power Flow}
The three-phase AC CC-OPF problem is formulated as follows:
\begin{subequations}
\label{eq:cc-form}
\begin{align}
      \min_{\substack{\boldsymbol{P}_{\mathbf{G},0}, {\boldsymbol{Q}}_\mathbf{G}, \\ |{\boldsymbol{V}}|, {\boldsymbol{\Theta}} }} ~&\sum_{\omega \in \Omega}\sum_{l \in \mathcal{N}}  {\text{VUF}}_{l,\omega}^2 &  \tag{CC-OPF} \\
      \text{s.t.} \qquad &  f\big(|{\boldsymbol{V}}_{i,\omega}|,\boldsymbol{\Theta}_{i,\omega},{{P}}_{i,\omega},{{Q}}_{i,\omega}\big)=0, \hspace{2.5em} \forall_{i \in \{0,\mathcal{N}\}, \omega\in\Omega}, \notag \\
      &\mathbb{P}_\omega \big( |{\boldsymbol{v}}_{i,\omega}^\phi| \leq \overline{v} \big) \geq 1 - \epsilon_{v},  \hspace{6.5em}\forall_{ \phi \in \Phi, i \in \mathcal{N}}, \notag\\
      &\mathbb{P}_\omega \big( |{\boldsymbol{v}}_{i,\omega}^\phi| \geq \underline{v} \big) \geq 1 - \epsilon_{v},  \hspace{6.5em}\forall_{\phi \in \Phi,i \in \mathcal{N}}, \notag\\
      &\mathbb{P}_\omega \!\big((\boldsymbol{q}_{\mathbf{G},i}^\phi)^2\! +\! (p_{\mathrm{G},i,\omega}^\phi)^2 \!\leq\! |s_{\mathrm{G},i}^\phi|\big)\!\geq\! 1-\epsilon_q,~ \forall_{\phi \in \Phi,i \in \mathcal{N}}, \notag\\
 &|{\boldsymbol{V}}_{0}|\angle {\boldsymbol{\Theta}}_{0}= {\begin{bmatrix}
        {1\angle 0^\circ} & {1\angle-120^\circ} & {1\angle 120^\circ}
    \end{bmatrix}}^\top,& \notag
\end{align}
\end{subequations}
where the optimization decision variables for each uncertainty realization~$\omega \in \Omega$ are the voltage magnitudes at all nodes~$|{\boldsymbol{V}}|\in \mathbb{R}^{3(n+1)}$, voltage angles at all nodes~${\boldsymbol{\Theta}}\in \mathbb{R}^{3(n+1)}$, active power generation at the substation~${\boldsymbol{P}}_{\mathbf{G},0}\in \mathbb{R}^{3}$, and reactive power generation at all nodes ${\boldsymbol{Q}}_\mathbf{G}\in \mathbb{R}^{3(n+1)}$. 

In general, we interpret the chance constraints~\eqref{eq:inv_pow} and \eqref{eq:Vlim} as the likelihood that the distribution grid operator will need to take extra control actions in real time to protect the system. By choosing high acceptable violation probabilities~$\epsilon_q \text{ and } \epsilon_v$, the system is at a higher risk of insecure operation as it may necessitate frequent deployment of real-time controls, which are not always available. Alternatively, choosing a low violation probability is expensive, but makes system operation safer and less stressful for the operator~\cite{roald2017chance}.

%% file: Sections/3_Solution_methods.tex
\section{Analytical Reformulation and Solution Methods}
\label{sec:solution-methods}
As formulated, CC-OPF is intractable due to the chance constraints~\eqref{eq:inv_pow} and \eqref{eq:Vlim}. Not only is the feasible region defined by the chance constraints non-convex, evaluating the feasibility of a solution is generally difficult as it requires multi-dimensional integration.
Furthermore, it can also be computationally expensive to enforce the power balance constraints~\eqref{eq:pow_bal} for all~$\omega \in \Omega$, particularly if the uncertainty set~$\Omega$ is large.
As a result, it is necessary to reformulate these constraints into deterministic counterparts in order to obtain a tractable formulation that can be efficiently solved. Our problem setting yields two additional complications. First, we do not assume any distributional assumptions on the uncertainty in the problem. Second, we must account for the uncertainty in the full, non-linear, non-convex AC power flow constraints. Consequently, CC-OPF falls under a challenging class of problems, where typical analytical reformulation methods cannot be applied.

\begin{figure}[!t]
	\centering	        	
	\includegraphics[width=0.5\textwidth]{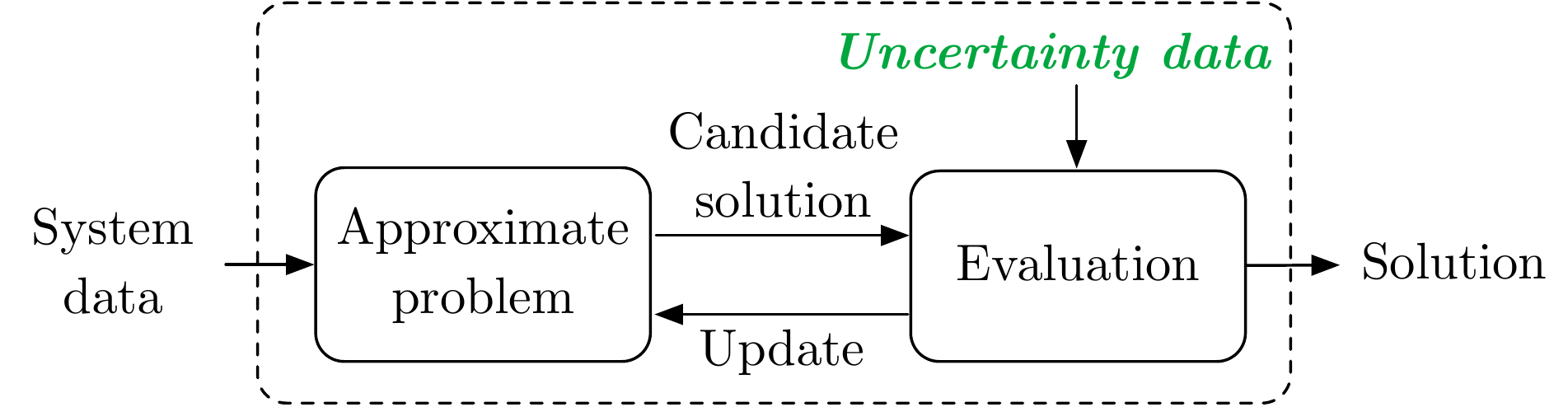}
	\caption{The general iterative process, consisting of solving an approximate problem formulation and using the results of a sample-based evaluation to update the formulation.}
	\label{fig:tuning-diagram}
\end{figure}

To solve this problem, we take an iterative, data-driven approach that combines an approximate problem formulation with sample-based evaluations to successively adapt the formulation, as illustrated in Fig. \ref{fig:tuning-diagram}. Inspired by~\cite{roald2017chance,hou2020chance}, we replace~\eqref{eq:pow_bal} by a single set of deterministic power flow equations and represent the chance constraints~\eqref{eq:inv_pow} and~\eqref{eq:Vlim} using their nominal counterpart plus a fixed tightening term that secures the system against uncertainty by functioning as an uncertainty margin. The tightening terms are successively updated using evaluated feedback from the available uncertainty data. 

The primary advantage of the iterative approach is that the solving of the three-phase AC OPF problem is \emph{decoupled} from the consideration of the uncertainty, which is wholly captured via the tightenings. This allows us to leverage the available uncertainty data without increasing the computational burden of solving the optimization problem. Thus, in contrast to other sample-based methods such as the scenario approach, the size of the optimization does not grow with the number of samples~\cite{hou2021data}.
Furthermore, we are able to use the full non-linear, non-convex AC power flow equations, without needing to perform any explicit linearizations or approximations. Because we deliberately choose to use an approximate reformulation that is simple to solve, the resulting algorithm can also take advantage of the computational efficiency of commercial solvers.

A challenge of using this style of deterministic reformulation is accurately determining and updating the values of the uncertainty margins. We need to ensure that the uncertainty margins are large enough that the solution to the reformulated problem is ensured to be feasible to the original chance constrained problem, yet small enough so that we do not obtain an overly conservative solution.
We thus explore two methods for iteratively updating constraint tightenings: a direct quantile-based method (inspired by~\cite{roald2017chance}) and a tuning-based method (based on~\cite{hou2020chance}). The remainder of this section details the main components of the iterative approach, including the two tuning variations.

\subsection{Approximate Problem Formulation}
\label{sec:reformulation}
As the approximate problem formulation, we choose to use a generalized reformulation based on~\cite{roald2017chance}. 
In the following, we detail the reformulated constraints and objective function.

\subsubsection{Power flow} Rather than enforcing the power flow equations in~\eqref{eq:pow_bal} for all uncertainty realizations, we enforce a single set of power flow equations at node~$i\in \{0,\mathcal{N}\}$ which are functions of the average power injections, $\bar{P}_{i},~\bar{Q}_{i}$ and corresponding voltage variables~$|\boldsymbol{V}_i|,\boldsymbol{\Theta}_i$. The shorthand representation of the power balance constraint becomes
\begin{align}
    & f(|\boldsymbol{V}_{i}|,\boldsymbol{\Theta}_{i},{\overline{P}}_{i},{\overline{Q}}_{i})=0, & \forall i \in \{0,\mathcal{N}\}. \notag
\end{align}

\subsubsection{Inverter limits}
\label{sec:det-q-limit}
\begin{figure}[t]
	\centering	        	
	\includegraphics[width=0.45\textwidth]{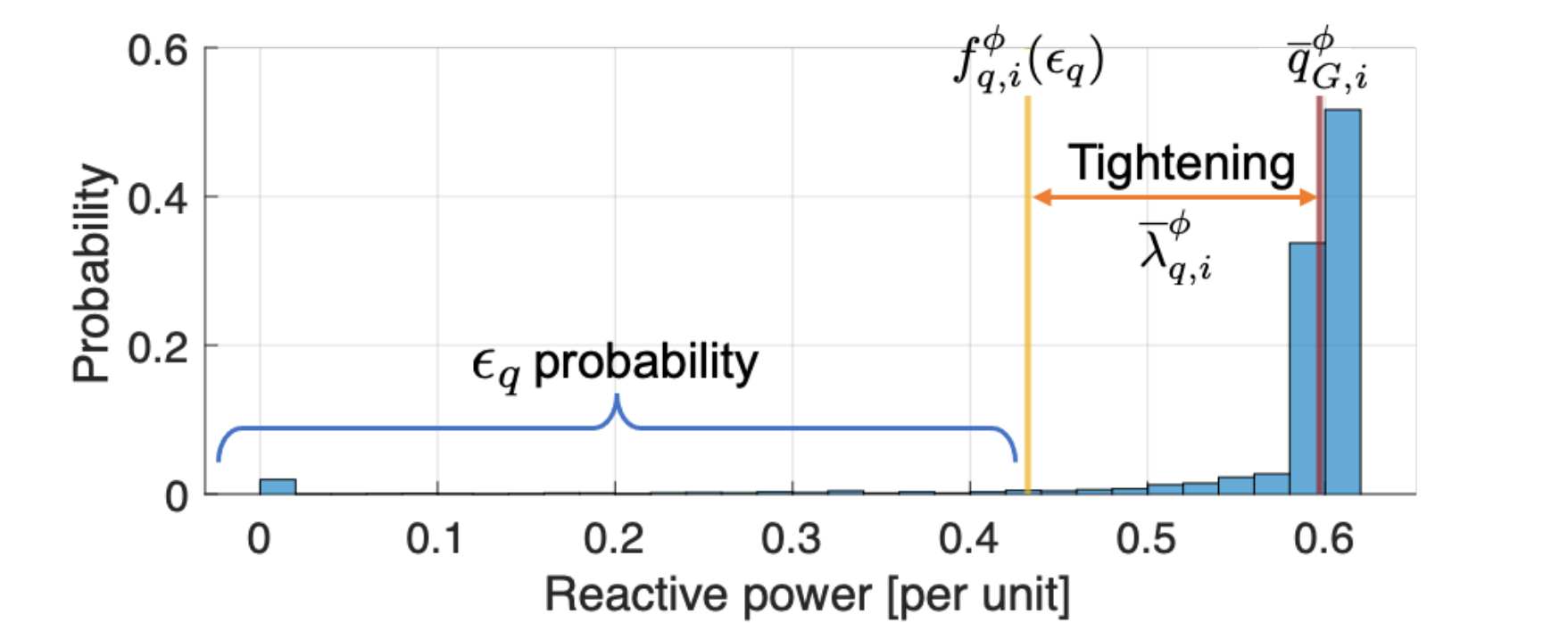}
	\caption{The histogram represents the probability distribution function of an example inverter reactive power upper limit, $f_{q,i}^{\phi}(\cdot)$. The constraint tightening $\overline{\lambda}_{q,i}^{\phi}$, illustrated with the orange line, is given by the distance between the nominal limit~$\overline{q}_{\mathrm{G},i}^{\phi}$ (brown line) and the~$\epsilon_q$ quantile of the empirical distribution of the upper reactive power limit~$f_{q,i}^{\phi}(\epsilon_q)$ (yellow line).}
	\label{fig:quantile-diagram}
\end{figure}

We replace the inverter limits defined by quadratic chance constraints in~\eqref{eq:inv_pow} with deterministic box constraints, consisting of a nominal limit with a tightening term. The upper and lower limits for a PV inverter connected to phase~$\phi \in \Phi$ at node~$i \in \mathcal{N}$ are calculated using the average generation $\overline{p}_{\text{G},i}^\phi$ as follows:
\begin{subequations}
\begin{align}
    \overline{q}_{\text{G},i}^{\phi} & = \sqrt{|{s_{\mathrm{G},i}^\phi}|^2-(\overline{p}_{\text{G},i}^\phi)^2}, &\forall \phi \in \Phi,i \in \mathcal{N}, \label{eq:q-upper} \\
    \underline{q}_{\text{G},i}^{\phi} & = -\sqrt{|{s_{\mathrm{G},i}^\phi}|^2-(\overline{p}_{\text{G},i}^\phi)^2}, &\forall \phi \in \Phi,i \in \mathcal{N}.
\end{align}
\end{subequations}
The tightenings for the upper and lower limits are denoted by $\overline{\lambda}_{q,i}^{\phi}, \underline{\lambda}_{q,i}^{\phi} \in \mathbb{R}^+$, respectively. For notational convenience, we also denote the tightenings using vectors
\begin{align*}
    &\underline{\Lambda}_{q} = \Big[\big[\underline{\lambda}_{q,i}^{\phi}\big]^{\top}_{\phi \in \Phi} \Big]_{i \in \mathcal{N} }^\top,~ &
    \overline{\Lambda}_{q} = \Big[\big[\overline{\lambda}_{q,i}^{\phi}\big]^{\top}_{\phi \in \Phi} \Big]_{i \in \mathcal{N} }^\top.
\end{align*}
Note that these tightenings can be pre-calculated since the inverter reactive power limits depend only on the uncertainty samples which are known a priori. E.g., for the upper inverter limit, we evaluate the limit $\overline{q}_{\text{G},i, \omega}^{\phi}$ for each uncertainty sample $\omega \in \Omega$ using Eq.~\eqref{eq:q-upper} with the sampled value $p_{\text{G}, i, \omega}$ rather than the averaged value $\overline{p}_{\text{G},i}^\phi$. By calculating this limit under all uncertainty realizations, we obtain an empirical distribution for the upper reactive power generation limit, which we denote as $f_{q,i}^\phi(\cdot)$. The empirical distribution for the lower reactive power generation limit would simply be the negative, i.e., $-f_{q,i}^\phi(\cdot)$. We then find the desired quantiles of the empirical distribution and use them to directly calculate the constraint tightenings. For the chance constraint on the upper reactive power limit to hold, we require the upper limit to be set to $f_{q,i}^{\phi}(\epsilon_q)$, which is the $\epsilon_q$ quantile of the empirical distribution of the upper reactive power limit. We can observe that the appropriate uncertainty margin to ensure the chance constraint holds corresponds to the difference between the nominal inverter lower limit ~$\underline{q}_{\mathrm{G},i}^{\phi}$ and~$f_{q,i}^{\phi}(\epsilon_q)$. Thus, the constraint tightenings for reactive power limits are calculated as follows:
\begin{align*}
    \overline{\lambda}_{q,i}^{\phi}  &= \overline{q}_{\mathrm{G},i}^{\phi}  - f_{q,i}^{\phi}(\epsilon_q), & \forall \phi \in \Phi, i \in \mathcal{N},\\
    \underline{\lambda}_{q,i}^{\phi}  &= -f_{q,i}^{\phi}(1 - \epsilon_q) - \underline{q}_{\mathrm{G},i}^{\phi}, & \forall \phi \in \Phi, i \in \mathcal{N}.
\end{align*}
This process is also illustrated in Fig. \ref{fig:quantile-diagram}. We note that once these tightening values are calculated, they remain fixed throughout the entirety of the iterative algorithm.

\subsubsection{Voltage limits} We similarly replace the chance constraints on the upper and lower voltage magnitude constraints ~\eqref{eq:Vlim} with the corresponding nominal limits and tightenings, defined by $\overline{\lambda}_{v,i}^{\phi}, \underline{\lambda}_{v,i}^{\phi} \in \mathbb{R}^+$, respcetively. The vector representations are
\begin{align*}
    &\underline{\lambda}_{V} = \left[\left[\underline{\lambda}_{V,i}^{\phi \in \Phi}\right]_{i \in \mathcal{N} }\right]^\top,~ 
    \overline{\lambda}_{V} = \left[\left[\overline{\lambda}_{V,i}^{\phi \in \Phi}\right]_{i \in \mathcal{N} }\right]^\top. \notag \\
\end{align*}
In contrast to the inverter limits, the voltage constraint tightenings need to be calculated and updated within the iterative algorithm since they require obtaining a solution from the approximate OPF problem in order to be determined.

\subsubsection{Objective function} We calculate the deterministic VUF by replacing the voltage variables~$\boldsymbol{V}_{l,\omega},\boldsymbol{\Theta}_{l,\omega}$ with their nominal counterparts~$|{\boldsymbol{V}_l}|, {\boldsymbol{\Theta}_l}$ (corresponding to average power injections) in~\eqref{eq:VUF_obj} and \eqref{eq:Vn_Vp}.

\subsubsection{Reformulated problem} The resulting approximate optimization problem obtained from using the reformulation described above is deterministic and takes the following form:
\begin{subequations}
\label{eq:cc-reform}
\begin{align}
      \min_{ \substack{{\boldsymbol{P}}_{\mathbf{G},0}, {\boldsymbol{Q}}_{\mathbf{G}}, \\ |{\boldsymbol{V}}|, {\boldsymbol{\Theta}}} } ~ & \sum_{l \in \mathcal{N}}  \text{VUF}_l^2 & & \tag{CCR-OPF} \\
      \text{s.t.} \qquad &  f(|{\boldsymbol{V}}_i|,{\boldsymbol{\Theta}_i},{\bar{P}}_{i},{\bar{Q}}_{i})=0,  & \forall i \in \{0,\mathcal{N}\}, & \notag \\
      & |{\boldsymbol{v}}_i^\phi| \leq \overline{v} - \overline{\lambda}_{v,i}^{\phi}, & \forall \phi \in \Phi, i \in \mathcal{N}, & \notag \\
      & |{\boldsymbol{v}}_i^\phi| \geq \underline{v} + \underline{\lambda}_{v,i}^{\phi}, & \forall \phi \in \Phi, i \in \mathcal{N}, & \notag \\
      & {{\boldsymbol{q}}}_{\mathbf{G},i}^\phi \leq \overline{q}_{\mathrm{G},i}^{\phi}-\overline{\lambda}_{q,i}^{\phi}, & \forall \phi \in \Phi, i \in \mathcal{N}, & \notag \\
      & {{\boldsymbol{q}}}_{\mathbf{G},i}^\phi \geq \underline{q}_{\mathrm{G},i}^{\phi}+\underline{\lambda}_{q,i}^{\phi}, & \forall \phi \in \Phi, i \in \mathcal{N}, & \notag \\
      & \boldsymbol{V}_0\angle \boldsymbol{\Theta}_0 = & & \hspace{-4.7cm} {\begin{bmatrix}
            {1\angle 0^\circ} & {1\angle-120^\circ} &  {1\angle 120^\circ}
        \end{bmatrix}}^\top. \notag
\end{align}
\end{subequations}
We denote a solution of CCR-OPF using $\mathbf{X} = ({\mathbf{P}}_{\mathbf{G},0}, {\mathbf{Q}}_{\mathbf{G}}, |{\mathbf{V}}|, {\boldsymbol{\Theta}})$.

\subsection{Solution Evaluation}
\label{sec:evaluation}
After we solve the approximate problem (CCR-OPF) and obtain a candidate solution $\mathbf{X}$, we use the available uncertainty samples to evaluate the conservativeness of the voltage constraint tightenings. We can either evaluate the solution $\mathbf{X}$ by creating an empirical distribution and performing a quantile evaluation (as done to obtain the inverter limit tightenings in Section \ref{sec:det-q-limit}) or by estimating the empirical violation probability of the solution. Both methods are detailed below.

\subsubsection{Quantile evaluation} 
\label{sec:quantile-eval}
We calculate power flows for each sample in our uncertainty set using the candidate solution $\mathbf{X}$. By doing so, we obtain empirical distributions for the voltage magnitude around the candidate voltage solution~$|{\mathbf{v}}_i^{\phi}|$ at node~$i \in \mathcal{N}$ connected to phase~$\phi \in \Phi$, which we denote with~$f_{v,i}^{\phi}(\cdot)$. Similarly to the evaluation of the inverter limit tightenings, we use the empirical distribution to directly evaluate the voltage magnitude value at our desired quantiles. As a result, we obtain $f_{v,i}^{\phi}(1-\epsilon_v)$ and $f_{v,i}^{\phi}(\epsilon_v)$ as the respective upper $1- \epsilon_v$ and lower $\epsilon_v$ quantiles of the empirical distribution~$f_{v,i}^{\phi}(\cdot)$.

\subsubsection{Empirical violation probability evaluation}
\label{sec:viol-prob-eval}
Alternatively, we can use the uncertainty samples to estimate the empirical violation probability of the candidate solution $\mathbf{X}$ to ascertain whether it satisfies the chance constraint. As an example, let us take the upper voltage magnitude constraint for node $i \in \mathcal{N}$ connected to phase $\phi \in \Phi$ and samples ${p}_{\text{G},i, \omega}^\phi$ and ${p}_{\text{L},i, \omega}^\phi$. We define an indicator random variable $Y_{\overline{v}, i}^{\phi}(\mathbf{X}, {p}_{\text{G},i, \omega}^\phi, {p}_{\text{L},i, \omega}^\phi) \in \{0,1\}$, which evaluates to 0 if the upper voltage constraint holds and 1 otherwise, i.e., 
\begin{align}
    Y_{\overline{v}, i}^{\phi}(\mathbf{X}, {p}_{\text{G},i, \omega}^\phi, {p}_{\text{L},i, \omega}^\phi) =
    \begin{cases}
        0 & \text{ if } |\boldsymbol{v}_{i,\omega}^{\phi}| \leq \overline{v}, \\
        1 & \text{ otherwise.}
    \end{cases}
\end{align}
We obtain the empirical violation probability for this constraint, denoted  $\hat{E}_{\overline{v},i}^{\phi}(\mathbf{X}) \in [0,1]$, by taking the average evaluated on all samples, i.e.,
\begin{align}
    \hat{E}_{\overline{v},i}^{\phi}(\mathbf{X}) = \frac{1}{|\Omega|} \sum_{\omega \in \Omega} Y_{\overline{v}, i}^{\phi}(\mathbf{X}, {p}_{\text{G},i, \omega}^\phi, {p}_{\text{L},i, \omega}^\phi).
\end{align}
We perform similar evaluations to obtain the empirical violation probabilities of the voltage lower limits~$\hat{E}_{\underline{v},i}^{\phi}(\mathbf{X})$, inverter upper limits~$\hat{E}_{\overline{q},i}^{\phi}(\mathbf{X})$, and inverter lower limits~$\hat{E}_{\underline{q},i}^{\phi}(\mathbf{X})$. For a solution $\mathbf{X}$, we define the worst case empirical violation probability for a single constraint type as the maximum empirical violation probability observed across all constraints of that type. For, e.g., the worst case empirical violation probability for the voltage upper limit is
\begin{align}
    \label{eq:wc-prob}
    \hat{E}^{\max}_{\overline{v}}(\mathbf{X}) = \max_{i \in \mathcal{N}, \phi \in \Phi} \{ \hat{E}_{\overline{v},i}^\phi \}.
\end{align}
We further define the worst case empirical across all voltage constraints as
\begin{align}
    \label{eq:wc-prob-all-v}
    \hat{E}_{v}^{\max}(\mathbf{X}) = \max \{ \hat{E}^{\max}_{\overline{v}}(\mathbf{X}), \hat{E}^{\max}_{\underline{v}}(\mathbf{X}) \}.
\end{align}

\subsubsection{Handling of Inverter Limits} 
As noted in Section \ref{sec:inverter-control}, we assume PV inverter limits are hard constraints. Consequently, as we evaluate the solution $\mathbf{X}$ in this step, we propose two methods to deal with the situation where the inverter limit specified in~\eqref{eq:inv_pow} is violated:
\begin{enumerate}[(i)]
    \item We allow inverters to be overloaded and observe violation of the inverter limits.
    \item We cap the inverter reactive power to be within the apparent power limit.
\end{enumerate}
We note that the latter approach (i.e., with capping) is a more realistic scenario. While the capping of the inverter output is difficult to represent as a constraint within the optimization problem itself, we can easily perform this a posteriori during the solution evaluation step.

\subsection{Iterative Schemes}
We next describe the two iterative algorithms used to solve the deterministic CCR-OPF.

\subsubsection{Quantile-based method}
\label{subsubsec:mc-method}
Based on~\cite{schmidli2016stochastic, roald2017chance, karagiannopoulos2017operational}, this method uses available samples to generate empirical distribution functions for the voltage variables. We then follow Section \ref{sec:quantile-eval} to find the desired quantiles of the empirical distribution, which are used to calculate the tightenings for the voltage magnitude limits. The quantile-based method comprises of the following steps:
\begin{enumerate}[(1)]
\item \emph{Initialize:} Set the iteration count to~$\kappa=0$~and voltage magnitude tightenings $\overline{\Lambda}_{v}^{(0)},\underline{\Lambda}_{v}^{(0)}$ to zero. The reactive power generation tightenings $\overline{\Lambda}_{q},\underline{\Lambda}_{q}$ are calculated  according to the procedure described in Section \ref{sec:det-q-limit}.

\item \emph{Solve approximate problem:} Solve CCR-OPF using fixed tightenings $\overline{\Lambda}_{v}^{(\kappa)},\underline{\Lambda}_{v}^{(\kappa)},\overline{\Lambda}_{q},\underline{\Lambda}_{q}$ to obtain a candidate operating point $\mathbf{X}^{(\kappa + 1)}$.

\item \emph{Solution evaluation:} Perform the quantile evaluation described in Section \ref{sec:quantile-eval} to obtain the upper $1 - \epsilon_v$ and lower $\epsilon_v$ quantiles of the empirical distribution of the voltage magnitude, $f_{v,i}^{\phi,(\kappa+1)}(1-\epsilon_v)$ and $f_{v,i}^{\phi,(\kappa+1)}(\epsilon_v)$.

\item \emph{Update tightenings:}
Use the evaluated quantiles to directly update the voltage magnitude constraint tightenings, 
\begin{align*}
    \overline{\lambda}_{v,i}^{\phi}  &= f_{v,i}^{\phi,(\kappa+1)}(1-\epsilon_v) - |{\boldsymbol{v}}_i^{\phi,(\kappa+1)}|, & \forall \phi \in \Phi,\forall i \in \mathcal{N}, \\
    \underline{\lambda}_{v,i}^{\phi}  &= |{\boldsymbol{v}}_i^{\phi,(\kappa+1)}| - f_{v,i}^{\phi,(\kappa+1)}(\epsilon_v), & \forall \phi \in \Phi,\forall i \in \mathcal{N}.
\end{align*}

\item \emph{Check convergence:} Terminate when the upper and lower voltage constraint tightenings converge below their respective pre-specified tolerance levels~$\overline{\eta}_v,\underline{\eta}_v \in \mathbb{R}^+$, i.e., both
\begin{align*}
    \max \{|\overline{\Lambda}_v^{(\kappa+1)}\!\!-\!\overline{\Lambda}_v^{(\kappa)}|\}  \leq \overline{\eta}_v, ~
    \max \{|\underline{\Lambda}_v^{(\kappa+1)}\!\!-\!\underline{\Lambda}_v^{(\kappa)}|\}  \leq \underline{\eta}_v
\end{align*}
are satisfied. Return final solution~$\mathbf{X}^{(\kappa + 1)}$. Otherwise, increase the iteration count to $\kappa = \kappa + 1$ and return to step (2).
\end{enumerate}

\subsubsection{Tuning-based method}
This solution algorithm employs a tuning-based approached adapted from~\cite{hou2020chance}. Rather than directly using the quantile evaluation to calculate the uncertainty margins, we define the tightenings as parameterized by a single-dimensional tuning parameter, denoted $s \in \mathbb{R}^{+}$. This parameter is then iteratively adjusted based on the empirical violation probability evaluation of the current candidate solution. We follow~\cite{hou2020chance} and use a simple bisection search to adjust $s$ (although we note that various other tuning procedures can also be used).

Before detailing the algorithm steps, we first define the tightening as a product between the tuning parameter $s$ and an approximation or estimation of the standard deviation of the nominal constraint. The tightenings are symmetric for the upper and lower voltage magnitude constraints and take the following form:
\begin{align}
    & \overline{\lambda}_{v,i}^{\phi} = \underline{\lambda}_{v,i}^{\phi} = s \cdot \sigma_{v,i}^\phi, &\forall \phi \in \Phi, i \in \mathcal{N}, \label{eq:Vlim_tm}
\end{align}
where $\sigma_{v,i}^\phi \in \mathbb{R}^{+}$ represents the estimated standard deviation of the voltage magnitude at node $i \in \mathcal{N}$ connected to phase $\phi \in \Phi$. 
To obtain this estimate, we solve CCR-OPF using voltage limit tightenings $\overline{\Lambda}_{v}=\underline{\Lambda}_{v}=0$ and inverter limit tightenings $\overline{\Lambda}_{q},\underline{\Lambda}_{q}$ initialized according to Section \ref{sec:det-q-limit}. We obtain an operating point $\mathbf{X}^{(0)}$ and evaluate the empirical distributions $f_{v,i}^{\phi,(0)}(\cdot)$ according to Section \ref{sec:viol-prob-eval}. We set $\sigma_{v,i}^\phi$ as the standard deviation of $f_{v,i}^{\phi,(0)}(\cdot)$.


The operating point $\mathbf{X}^{(0)}$ is also used to initialize upper bound~$s_{\max}^{(0)}$ and lower bound~$s_{\min}^{(0)}$ of the tuning parameter. We aim to initialize these bounds such that the mid-point of the bounds (which will be the initial tuning parameter value $s^{(0)}$ since we are performing a bisection search) will approximate the initial tightenings of the quantile-based method. The initial bounds are as follows:
\begin{subequations}
\label{eq:s-bound-init}
\begin{align}
    s_{\min}^{(0)} &= 0, \\
    s_{\max}^{(0)} &= \Big( \max_{\substack{i \in \mathcal{N}, \phi \in \Phi}} \big\{ |\mathbf{v}_i^{\phi, (0)}| - f_{v,i}^{\phi, (0)}(\epsilon_v) \big\} \Big) \frac{2}{\sigma_{v,i}^{\phi}}.
\end{align}
\end{subequations}
The lower bound corresponds to the case where all tightenings are zero.
The upper bound calculates the maximum difference between the voltage magnitude operating point and $\epsilon_v$ quantile across all single phase connections. The resulting value is divided by the standard deviation estimate and doubled.

\begin{enumerate}[(1)]

\item \emph{Initialize:} Set the iteration count to $\kappa=0$ and calculate the inverter limit tightenings~$\overline{\Lambda}_{q},\underline{\Lambda}_{q}$ as described in Section \ref{sec:det-q-limit}. Initialize the tuning parameter bounds $s_{\min}^{(0)},s_{\max}^{(0)}$ according to \eqref{eq:s-bound-init} and calculate the initial value of the tuning parameter by taking the mid-point of the bounds $s^{(0)} = (s_{\max}^{(0)} - s_{\min}^{(0)})/2 + s_{\min}^{(0)}$. Calculate $\overline{\Lambda}_{v}^{(0)}, \underline{\Lambda}_{v}^{(0)}$ according to~\eqref{eq:Vlim_tm}.

\item \emph{Solve approximate problem:} Solve CCR-OPF using fixed tightenings $\overline{\Lambda}_{v}^{(\kappa)},\underline{\Lambda}_{v}^{(\kappa)},\overline{\Lambda}_{q},\underline{\Lambda}_{q}$ to obtain  candidate operating point $\mathbf{X}^{(\kappa + 1)}$.

\item \emph{Solution evaluation:} Perform the empirical violation probability evaluation described in Section \ref{sec:viol-prob-eval} to calculate the worst case empirical violation probability $\hat{E}^{\max}_{v}(\mathbf{X})$ as defined in \eqref{eq:wc-prob-all-v}.

\item \emph{Update tightenings:}\\
If $\hat{E}^{\max}_{v}(\mathbf{X}^{(\kappa+1)}) \leq \epsilon_v$, decrease $s$ by setting $s_{\max}^{(\kappa+1)} = s^{(\kappa)}$ to obtain a less conservative solution.\\
If $\hat{E}_{v}^{\max}(\mathbf{X}^{(\kappa+1)}) > \epsilon_v$, increase $s$ by setting $s_{\text{min}}^{(\kappa+1)} = s^{(\kappa)}$ to obtain a more conservative solution.\\ Obtain a new tuning parameter $s^{(\kappa + 1)} = (s_{\max}^{(\kappa + 1)} - s_{\min}^{(\kappa + 1)})/2 + s_{\min}^{(\kappa + 1)}$ and update the  tightenings according to~\eqref{eq:Vlim_tm}.

\item \emph{Check convergence:} Terminate if the worst case violation probability is within a tolerance level~$\eta \in \mathbb{R}^+$ of the desired $\epsilon_v$, i.e, $|\hat{E}^{\max}_{v}(\mathbf{X}^{(\kappa+1)}) - \epsilon_v | \leq \eta$, or if the upper and lower bounds have converged within tolerance $\eta_s \in \mathbb{R}^+$, i.e. $|s_{\text{max}}^{(\kappa)} - s_{\text{min}}^{(\kappa)}| \leq \eta_s$. 
Return the solution with the lowest objective that satisfies $|\hat{E}^{\max}_{v}(\mathbf{X}^{(\kappa+1)}) - \epsilon_v | \leq 0$.
Otherwise, increase iteration count $\kappa=\kappa+1$, and go back to step (2).
\end{enumerate}

\subsubsection{Comparison of iterative schemes}
\label{sec:alg-comparison}
It is evident that both quantile-based and tuning-based methods share large similarities, mainly differing in how the uncertainty margins are updated in each iterations. The most significant difference between the two methods is that the quantile-based method has more flexibility in its tightening updates due to the ability to calculate different tightenings for each of the voltage constraints separately. In contrast, the tuning method relies on adjusting a single tuning parameter $s$, which is used across all voltage constraints. However, the drawback of the flexibility of the quantile-based method is that there is a higher potential for overfitting to the sampled uncertainty data.

The updating of the tightenings of the two methods are loosely related. For the tuning-based method, the tightenings are defined as the product of $s$ and an estimate of the standard deviation. We can interpret tuning $s$ as choosing the number of standard deviations away from the average value to set the voltage limits. If the distributions $f_{v,i}^{\phi}$ were known, then we could choose the number of standard deviations to directly correspond to the $\epsilon_v$ or $1 - \epsilon_v$ quantiles of $f_{v,i}^{\phi}$. In other words, tuning essentially functions as an indirect method for choosing a quantile.


%% file: Sections/4_Case_study.tex
\section{Case Study Set-Up}
\label{sec:case-study}
In the case studies, we perform extensive numerical simulations on the IEEE 13-node radial distribution feeder~\cite{schneider2017analytic} using realistic uncertainty data for solar PV generation and load demand to assess the performance of the iterative algorithms detailed in Section \ref{sec:solution-methods}.
Our goal is to find reactive power set-points for the inverters that are valid for the whole day, i.e., we assume that the distribution of possible load and solar PV values is representative for power injection profiles across the whole day. 

We investigate how properties of the uncertainty data set (i.e., sampling procedure, number of samples, correlation between data points) can impact the resulting solutions. We further compare the performance of the two algorithms in terms of chance constraint feasibility and objective optimality using both in- and out-of-sample evaluations. In this section, we describe the test case and data sets used to obtain the results in the subsequent sections. 

The optimization problem and algorithms are implemented in Julia~\cite{julia}~using JuMP~\cite{DunningHuchetteLubin2017}~with the solver Ipopt~\cite{Ipopt}. 

\subsection{Feeder Description}
\begin{figure}[!t]
	\centering	        	
	\includegraphics[width=0.35\textwidth]{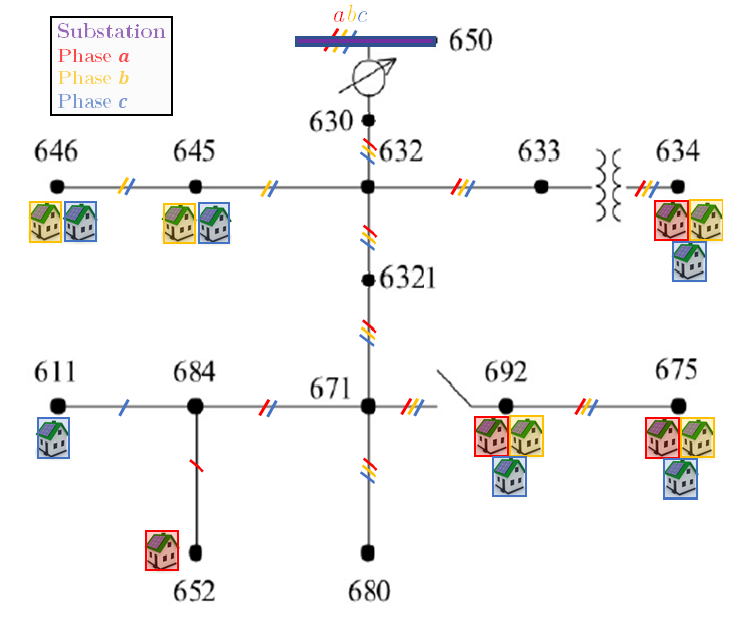}
	\caption{Modified IEEE-13 node feeder.}
	\label{fig:13feeder}
\end{figure}

As our test system, we use the modified IEEE 13-node feeder~\cite{schneider2017analytic}, which consists of 15 houses represented as single-phase connections at seven nodes as shown in Fig.~\ref{fig:13feeder}. 
To ensure that the generation and demand data is realistic, we use Pecan Street~\cite{street2015dataport} residential data with 1-minute resolution. This data comprises of measurements from residential homes in New York, which are available for 25 full days between May 2019 and August 2019. The measurements include rooftop solar PV generation and load profiles modelling electric vehicle charging behavior, HVAC, refrigerator, and other appliance use. Recall that an example of the load and solar PV profiles for a single day was shown in Fig.~\ref{fig:meas_data}. The Pecan Street data was scaled by a factor of 20 to match the existing loads in IEEE 13-node feeder. We assume each house has a single-phase rooftop solar PV system with a maximum inverter rating of 100kVA. 

\subsection{Sampling Procedure}
For our numerical simulations, we split the available data for in-sample and out-of-sample evaluations.

\subsubsection{Out-of-sample data}
We first randomly choose five days of data from the 25 day data set to be used for an out-of-sample evaluation. These samples are the same across all experiments.

\subsubsection{In-sample data}
\label{sec:sampling}
We draw~$M$ samples from the remaining 20 days to be used in the iterative algorithms and in-sample evaluation. These~$M$ samples are drawn using two types of sampling methods with three size variations:
\begin{enumerate}[(i)]
    \item \emph{Full day samples}: 
    We randomly choose either 1, 2 or 4 days from the set of 20 days, corresponding to $M = 1440,~ 2880,~\text{or}~ 5760$ data points, respectively. This data set corresponds to the common practice of using data from a set of ``representative days". In this case, there may be strong correlations among subsequent time steps, but all time steps will be represented in the data used in the algorithms.
    \item \emph{Random samples}: All data points from the set of 20 days are pooled together and~$M = 1440,~ 2880,~\text{or}~ 5760$ samples are randomly drawn from the data pool. This data set assumes that the data from the 20 days represents a probability distribution of the data and draws i.i.d. samples from this set. This does not guarantee that all time steps in the day are represented, but is likely to represent a wider variety of operating conditions as it includes data from more days.
\end{enumerate}

\subsection{Investigations} 
In the following two sections, we perform a variety of analyses to examine the performance of our methods and assess the different ways of modeling hard inverter limits:
\begin{itemize}
    \item Section \ref{sec:case-study-1} investigates how \textbf{different data sampling procedures} impact the performance of our methods. Specifically, we analyze the performance of the two iterative algorithms with the full day and random samples across several different algorithm replications, and aim to assess which sampling procedure is most suitable. 
    
    \item After determining the most suitable sampling approach, Section \ref{sec:single-rep-results} provides a \textbf{detailed comparison between the two iterative algorithms}. We look at the resulting constraint tightenings, inverter and voltage set points, and constraint violations evaluated on in- and out-of-sample data.
    
    \item Section \ref{sec:case-study-2} investigates algorithm performance when the effect of \textbf{inverter capping} is incorporated. 
\end{itemize}

\begin{figure*}[!t]
	\centering
	\includegraphics[width=0.90\textwidth]{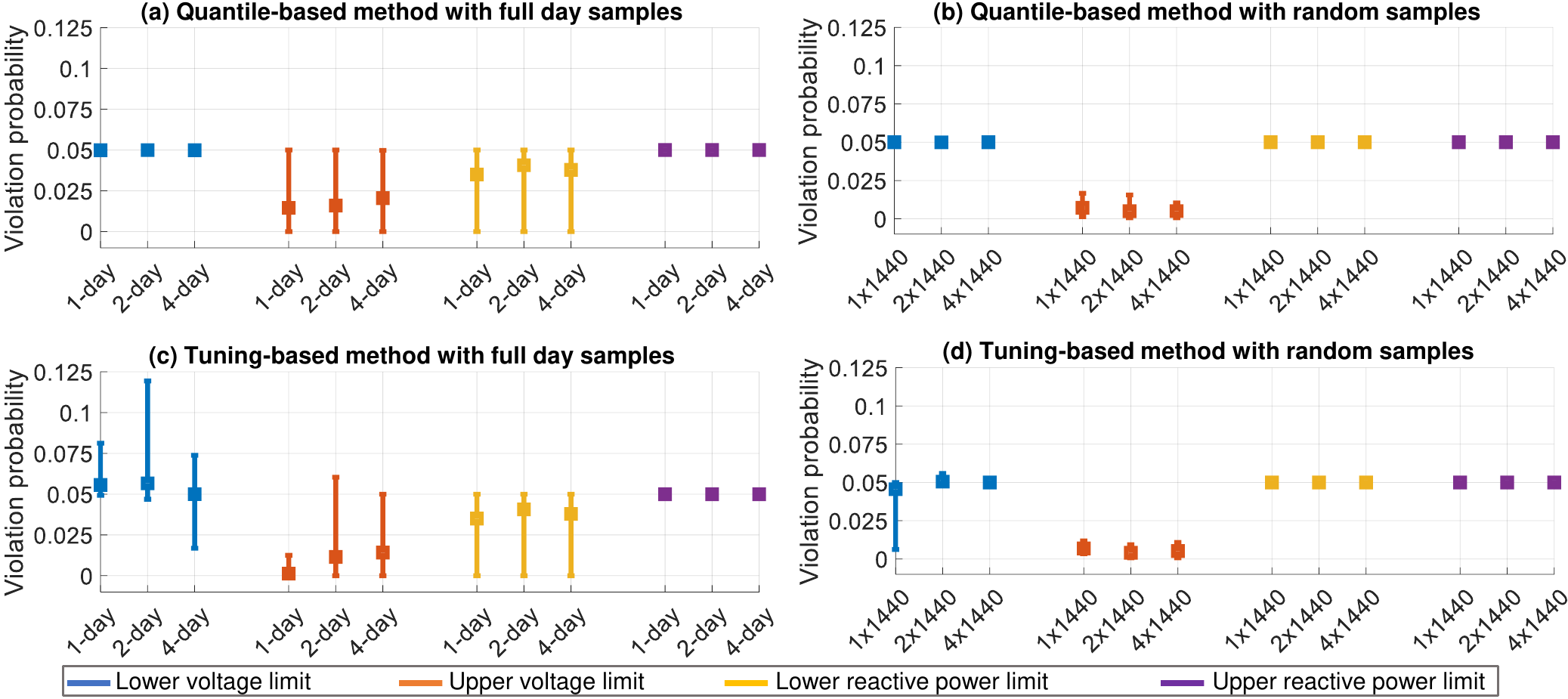}
	\caption{In-sample evaluation of violation probabilities for quantile-based and tuning-based  methods with~$1,2,4$ full day samples and~$1440,2880,5760$ random samples for 10 replications. The worst-case empirical violation probability for all constraint types are shown. The square represent the average values over 10 replications and the bars show the variability of violation probabilities (minimum and maximum values) across 10 replications.}
	\label{fig:QMvsTM_in-sample}
\end{figure*}

\section{Case Study I: Comparison of Sampling Methods}
\label{sec:case-study-1}
We first run numerical experiments to assess the performance of both the quantile and tuning-based methods under each of the data set sampling variations detailed in Section \ref{sec:sampling}. For each of the six sampling methods (full day or random samples with $M=1440,~ 2880,~\text{or}~ 5760$), we perform 10 algorithm replications, each using a different, independent sample draw. We set the desired chance constraint violation probability to be $\epsilon_v = \epsilon_q = 0.05$.

In Figs. \ref{fig:QMvsTM_in-sample} and \ref{fig:QMvsTM_out-sample}, we plot the worst-case empirical violation probabilities for each constraint type \eqref{eq:wc-prob} evaluated on in- and out-of-sample data, respectively. The worst-case violation probabilities for the lower voltage magnitude, upper voltage magnitude, lower inverter reactive power, and upper inverter reactive power constraints are illustrated in the figures using the blue, red, yellow, and purple bars, respectively. The average across the 10 replications is shown using the square, while the bars represent the range of worst-case violation probabilities obtained across replications. The figures illustrate the results for solutions obtained by the quantile-based (top) and tuning-based (bottom) methods, when using full day samples consisting of 1, 2, and 4 days (left) and $M = 1\times1440, 2\times1440$, and $4\times1440$ randomly sampled data points (right).

We assess solution optimality by calculating the total VUF obtained the CCR-OPF solutions. Fig.~\ref{fig:QMvsTM_in-sample_VUF} shows the average, minimum, and maximum values across 10 replications of the in-sample VUF for solutions obtained by each algorithm and data set variation. We also evaluate the solution quality on the out-of sample data by calculating the difference between the out-of-sample VUF and in-sample VUF normalized by the in-sample VUF. The out-of-sample VUF for each replication is obtained by calculating the total unbalance averaged across the five evaluation days. Fig.~\ref{fig:QMvsTM_out-sample_VUF} illustrates the average, minimum, and maximum values across 10 replications of the normalized VUF for each algorithm and data set variation, evaluated on out-of-sample data.

\subsection{Comparison of In-Sample Results}
We compare the in-sample results of both methods using the various data sets.

\subsubsection{Empirical violation probability} To assess the feasibility of solutions to the original chance-constrained problem, we compare the in-sample violation probablities of solutions obtained from the quantile-based method (top) and the tuning-based method (bottom) in Fig~\ref{fig:QMvsTM_in-sample}. We observe that the quantile-based method limits the worst-case violation probabilities of all constraint types to the desired $\epsilon_v = \epsilon_q = 0.05$ for all cases. The tuning-based method is unable to consistently obtain a worst case violation probability of $0.05$ using the full day data, particularly for the upper voltage constraints, but mostly achieves violation probabilities close to $0.05$ when using randomly sampled data. This may be a consequence of the tuning-based method having less flexibility than the quantile-based method in adapting the tightenings. While the quantile-based methods adjusts all tightenings individually based on the quantile evaluation, the tuning-based method uses the same tuning parameter $s$ across all constraints, including both upper and lower voltage magnitude constraints. As a result, the quantile-based method provides more ``fine-tuned'' and less conservative tightenings. 

\begin{figure*}[!b]
	\centering	        	
	\includegraphics[width=0.90\textwidth]{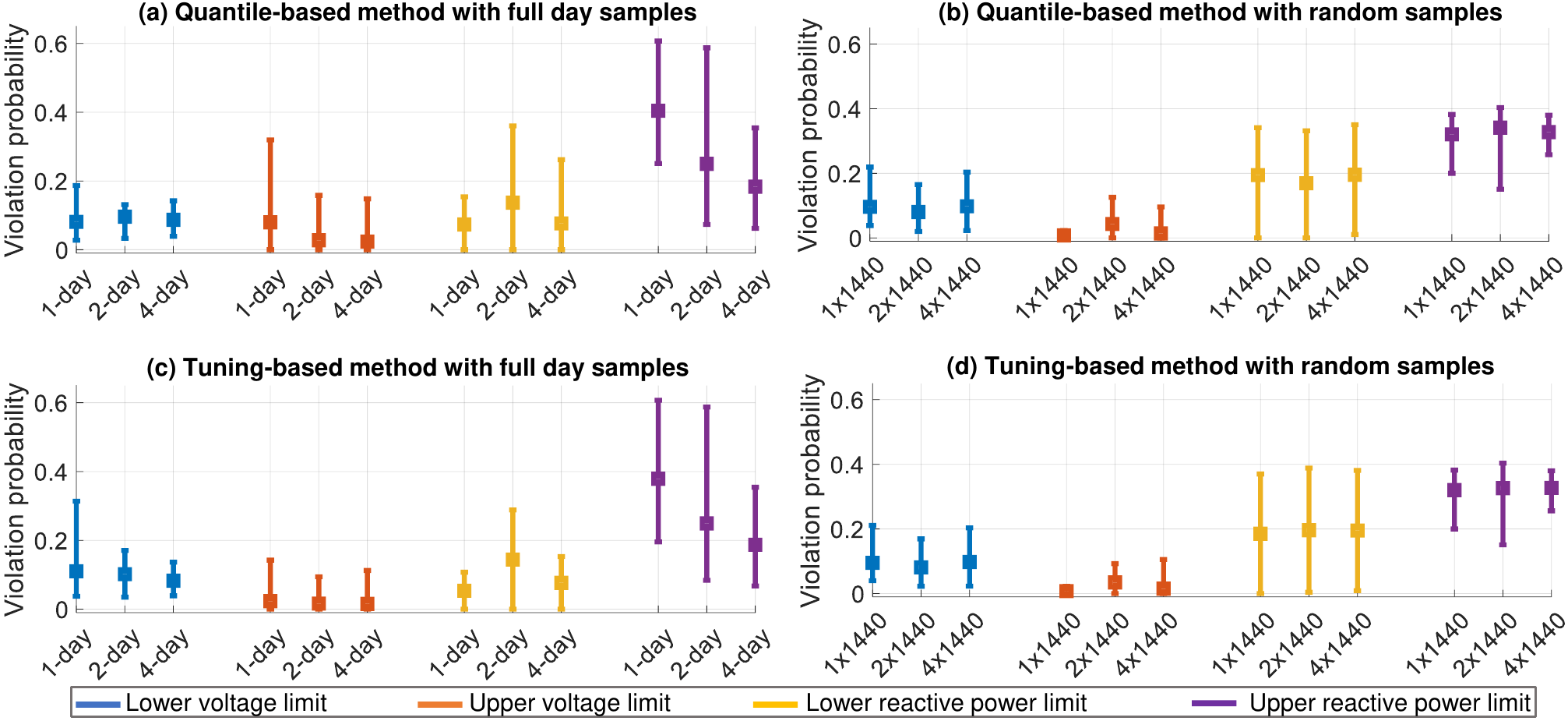}
	\caption{Out-of-sample evaluation of violation probabilities for quantile-based and tuning-based  methods with~$1,2,4$ full day samples and~$1440,2880,5760$ random samples for 10 replications. The worst-case empirical violation probability for all constraint types are shown. The square represent the average values over 10 replications and the bars show the variability of violation probabilities (minimum and maximum values) across 10 replications.}
	\label{fig:QMvsTM_out-sample}
\end{figure*}

\begin{figure}[!t]
	\centering	        	
	\includegraphics[width=0.45\textwidth]{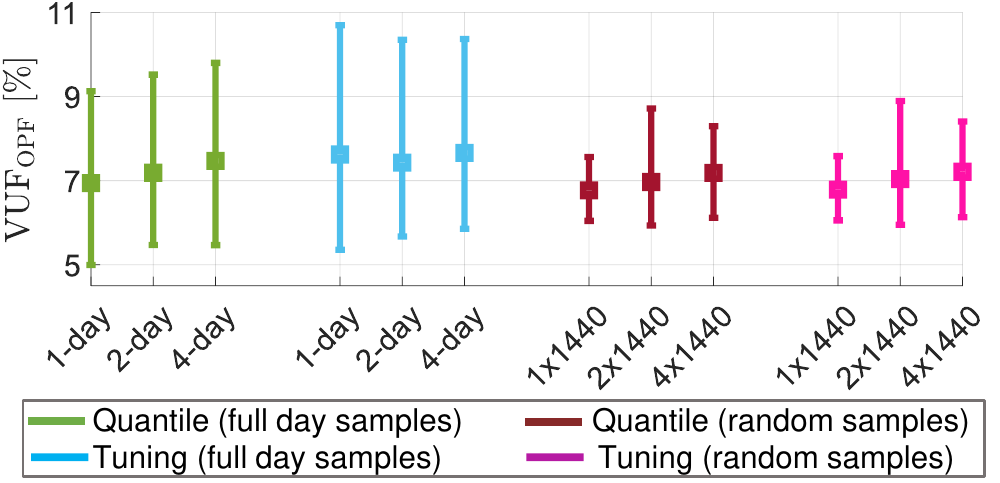}
	\caption{In-sample evaluation of total VUF for quantile-based  and tuning-based  methods with~$1,2,4$ full day samples and~$1440, 2880,5760$ random samples for 10 replications. The square represent the average values over 10 replications and the bars show the variability of VUF (minimum and maximum values) across 10 replications.}
	\label{fig:QMvsTM_in-sample_VUF}
\end{figure}

\subsubsection{Sampling procedure}
In Fig.~\ref{fig:QMvsTM_in-sample}, we observe that, for both algorithms, the range of worst-case probabilities is much smaller when using random samples (right) than when using full day samples (left). This likely occurs since drawing random samples results in a data set that captures a wider range of possible uncertainty realizations.

\subsubsection{VUF}
Due to the less conservative tightening approach of the quantile-based method, we observe in Fig.~\ref{fig:QMvsTM_in-sample_VUF} that the average VUF values for the quantile-based method (green and maroon bars) are slightly lower than the VUF obtained by the tuning-based method (light blue and magenta bars) which solves a more constrained problem resulting in higher unbalance levels. Furthermore, by comparing the results obtained with the full day samples (green and light blue bars) and the results obtained using random samples (maroon and magenta bars), we see that the VUF tends to be slightly smaller on average and fall in a narrower range. This is true across all sample sizes (including the smallest sample size set). This indicates that the total set of operating conditions is much better represented by using randomly chosen samples rather than using full days of data.

\subsection{Comparison of Out-of-Sample Results} 
\label{sec:oos-results}
We next consider the out-of-sample performance of each method.

\subsubsection{Empirical violation probability}
In Fig.~\ref{fig:QMvsTM_out-sample}, we first observe that the maximum violation probability across the 10 replications is significantly higher than the desired violation probabilities $\epsilon_v=\epsilon_q=0.05$, with empirical violation probabilities up to $0.6$. This indicates that solutions obtained from either method are likely not feasible to the original chance constrained problem. When comparing the quantile-based results (top) and the tuning-based results (bottom), we observe that both method generally provide comparable out-of-sample violation probabilities.

\subsubsection{Sampling procedure} The significant difference in the in- and out-of-sample violation probabilities additionally suggests that the data used within the optimization algorithm (i.e., to produce the in-sample results) is not representative of the out-of-sample data, contributing to the highly disappointing results. We do, however, observe that the solutions obtained using randomly selected samples (right) tend to have lower worst-case violation probabilities, indicating that randomly chosen samples may be an advantageous sampling procedure for both iterative algorithms. We further observe that the use of random samples results in solutions with smaller ranges (i.e., less variation across replications) of violation probabilities compared to full-day samples.

\subsubsection{VUF} 
\begin{figure}[!t]
	\centering	        	
	\includegraphics[width=0.45\textwidth]{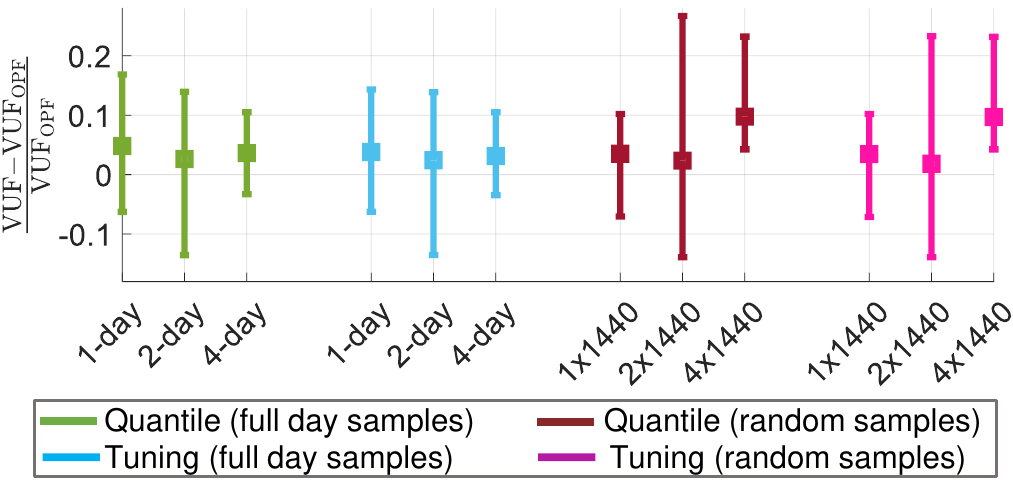}
	\caption{Out-of-sample evaluation of normalized VUF for quantile-based  and tuning-based  methods with~$1,2,4$ full day samples and~$1440, 2880,5760$ random samples for 10 replications. The square represent the average values over 10 replications and the bars show the variability of VUF (minimum and maximum values) across 10 replications.}
	\label{fig:QMvsTM_out-sample_VUF}
\end{figure}

From Fig.~\ref{fig:QMvsTM_out-sample_VUF}, we observe that there is negligible difference between the results obtained by the quantile-based method (green and maroon bars) and results for the tuning-based method (light blue and magenta bars). Furthermore, the deviation in out-of-sample VUF from the in-sample VUF is higher when we use random samples (maroon and magenta bars) compared to full day samples (green and light blue bars). This is likely because the full day in-sample data is more representative of the out-of-sample data, where we also use full day samples. 

Overall, we conclude that both methods perform in a similar manner, with the quantile-based method obtaining slightly better in-sample results. 

\section{Case Study II: Comparison of Iterative Algorithms}
\label{sec:single-rep-results}
The subsequent sections discuss results of a single replication of each of the iterative algorithms. Based on the conclusions of Section \ref{sec:oos-results}, we choose to use a data set comprising of $M = 2880$ randomly sampled data points.

\subsubsection{Constraint tightenings} 
\begin{figure}[!t]
	\centering	        	
	\includegraphics[width=0.32\textwidth]{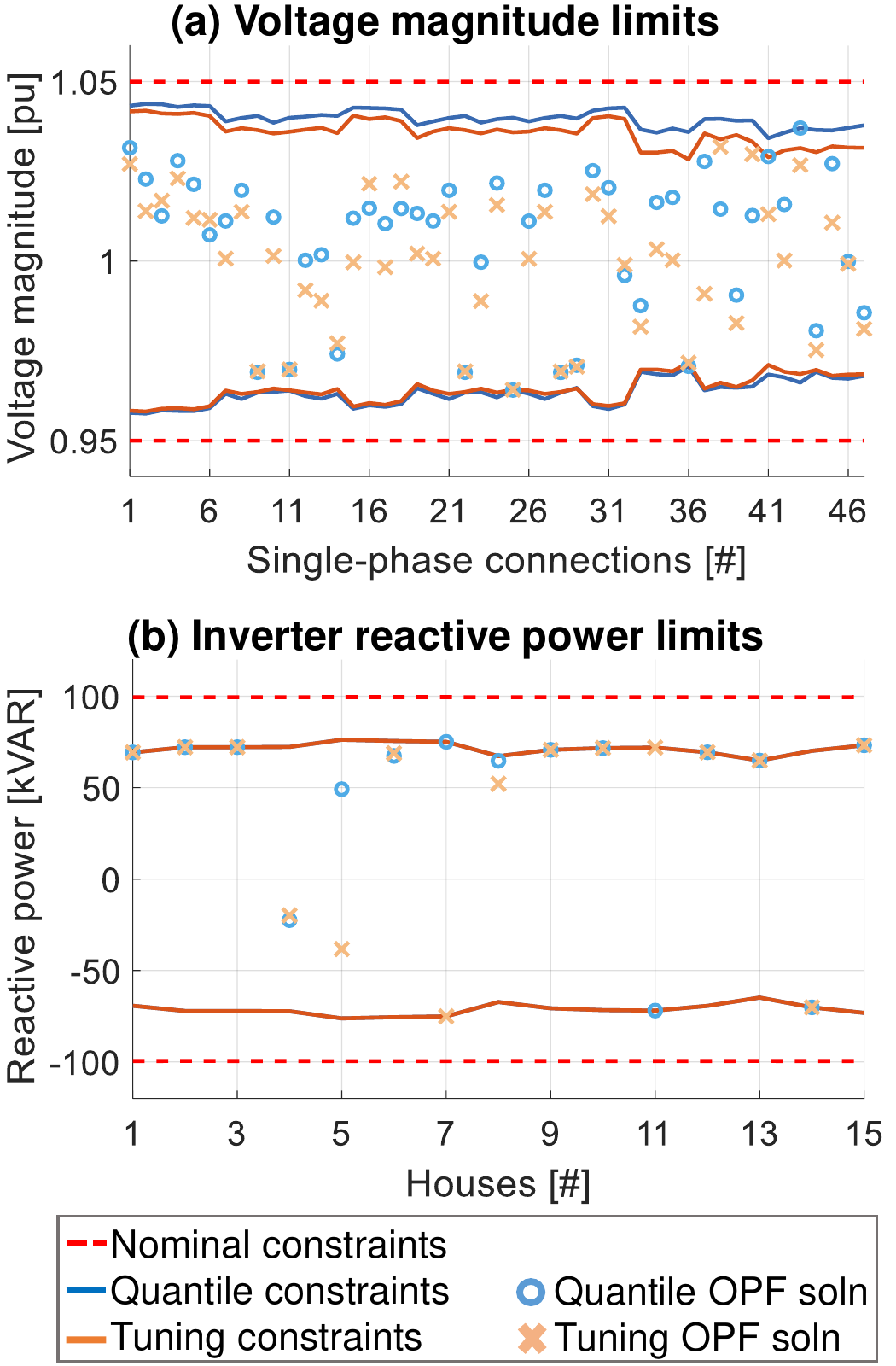}
	\caption{Comparison of CCR-OPF solution and constraint tightenings resulting from the quantile-based and tuning-based methods for single replication using~$M=2880$ random samples.}
	\label{fig:constr_trial5}
\end{figure}

We first compare the constraint tightenings obtained with the two different methods.
Shown in Fig.~\ref{fig:constr_trial5} are the voltage magnitude set-points and constraint tightenings across all single-phase connections (left) and PV inverter reactive power set-points and constraint tightenings across all houses (right). The nominal constraints are plotted in a dashed red line, while the tightened constraints are shown as blue lines (quantile-based method) and orange lines (tuning-based method). The tightening is the difference between the red dashed and the respective solid lines. We first observe that the inverter reactive power constraints for both methods shown in Fig.~\ref{fig:constr_trial5}(b) are identical. This is as expected because they are calculated in a similar manner as described in Section \ref{sec:det-q-limit}.

We consider the voltage magnitude constraints shown in Fig.~\ref{fig:constr_trial5}(a). We first observe that the tightenings for the upper and lower voltage constraints obtained with the tuning-based method (orange lines) are symmetric. This is as expected, because the voltage tightening of the tuning-based method is given by the product between the standard deviation of the voltage magnitude at each node (which is the same for the upper and lower bound) and the tuning parameter $s$ (which is shared among all voltage constraints) and is thus the same for both the upper and lower bound. 
We further notice that the tuning-based method is able to find the tuning parameter $s$ that results in a tightening of the lower voltage bound (orange line) that closely matches the tightening determined by the quantile-based method (blue line). This empirically demonstrates that tuning $s$ to be the correct number of standard deviations closely approximates the $\epsilon_v$ quantile for this constraints. On the other hand, the tightening of the upper voltage limits do not match as closely, i.e., there is a larger difference between the tightenings obtained with the tuning-based method (orange line) and the quantile-based method (blue line). This indicates that the probability distribution of the voltage magnitudes is not symmetric. This non-symmetry cannot be captured by the tuning-based method as explained above. 

\subsubsection{OPF solutions} 
As a result of the variations in the constraint tightenings, we obtain slightly different solutions with the two methods. The VUF of the nominal solution is~$6.3\%$ for the quantile-based method and~$6.5\%$ for the tuning-based method. This is as expected, since the quantile-based methods has lower constraint tightenings and thus a slightly larger feasible space, allowing for better solutions. 

The nominal voltage magnitude and reactive power obtained with each method are shown in Fig.~\ref{fig:constr_trial5}, with blue circles representing the quantile-based solutions and the yellow crosses representing the tuning-based solutions. When comparing the reactive power set-points obtained with the two methods in Fig.~\ref{fig:constr_trial5}(b), we see that many of the set-points are either the maximum or minimum allowable reactive power injections and that the two methods mostly produce similar reactive power set-points. However, there are also notable differences. For example, the inverters at houses 7 and 11 have high reactive power injection, but with opposite signs.
These differences in the reactive power set-points leads to differences in the voltage magnitude values, shown in Fig.~\ref{fig:constr_trial5}(a) with the quantile-based method producing generally higher voltage magnitudes than the tuning-based method. In particular, we observe that the quantile-based method produces a nominal voltage magnitude at node 43 that violates the tightened constraint of the tuning-based method. This shows how smaller constraint tightenings can enable solutions with a lower nominal VUF value. 

\subsubsection{Constraint violations by node and inverter} 
\begin{figure}[!b]
	\centering	        	
	\includegraphics[width=0.49\textwidth]{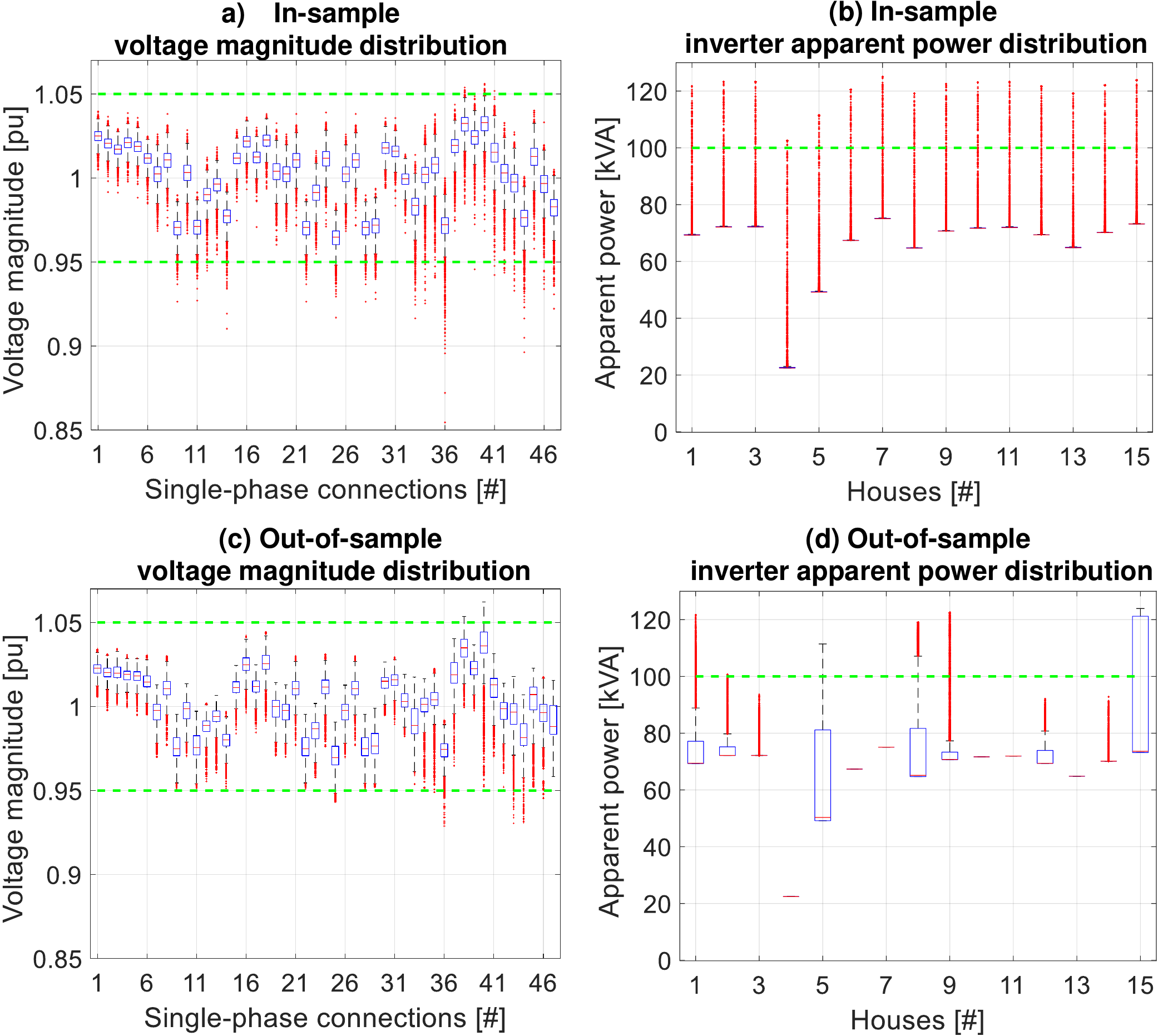}
	\caption{Box-whisker plots for voltage magnitude and inverter apparent power distribution calculated using Monte Carlo simulations for the quantile-based method for single replication using~$M=2880$ random samples. The green dashed lines represent the lower and upper voltage magnitude limits and the maximum inverter apparent power limits.}
	\label{fig:QM_distr_trial5}
\end{figure}

We next compare the performance of the methods in terms of constraint violations for the same replication discussed above. To assess the differences in performance, we compute the voltage magnitude and inverter apparent power for each in-sample and out-of-sample realization. Fig.~\ref{fig:QM_distr_trial5} shows the box-whisker plots of the results for the in-sample (top) and out-of-sample (bottom) data using the quantile-based solution. We omit the results obtained for the tuning-based method, as they are qualitatively similar to the ones seen in  Fig.~\ref{fig:QM_distr_trial5}.

We first compare the in-sample and out-of-sample results for the voltage magnitudes (left). 
The voltage magnitude violations in the in-sample evaluation (shown in Fig.~\ref{fig:QM_distr_trial5}(a)) are both larger in magnitude and occur at a greater number of nodes compared to the out-of-sample evaluation (shown in Fig.~\ref{fig:QM_distr_trial5}(c)). A similar trend is observed when we compare the in-sample and out-of-sample constraint violations of the inverters. While the largest constraint violations are of similar magnitudes in both the in-sample and out-of-sample data, the number of nodes experiencing violations is much larger in the in-sample evaluation. 
Moreover, we see that the variability in the data for both voltage magnitudes and apparent power is larger in the in-sample compared to the out-of-sample distributions. 
This further suggests that the in-sample data is not representative of the out-of-sample distributions and may be due to the difference in the sampling method for the in-sample evaluation, where random samples are used, and the out-of-sample evaluation, where 5 full days of data is used.
Furthermore, this indicates the necessity of ensuring the days used for the out-of-sample evaluation are representative of actual operating conditions.

\subsubsection{Constraint violations by time of day} 
\label{sec:single-rep-time-series}
\begin{figure}[!b]
	\centering	        	
	\includegraphics[width=0.49\textwidth]{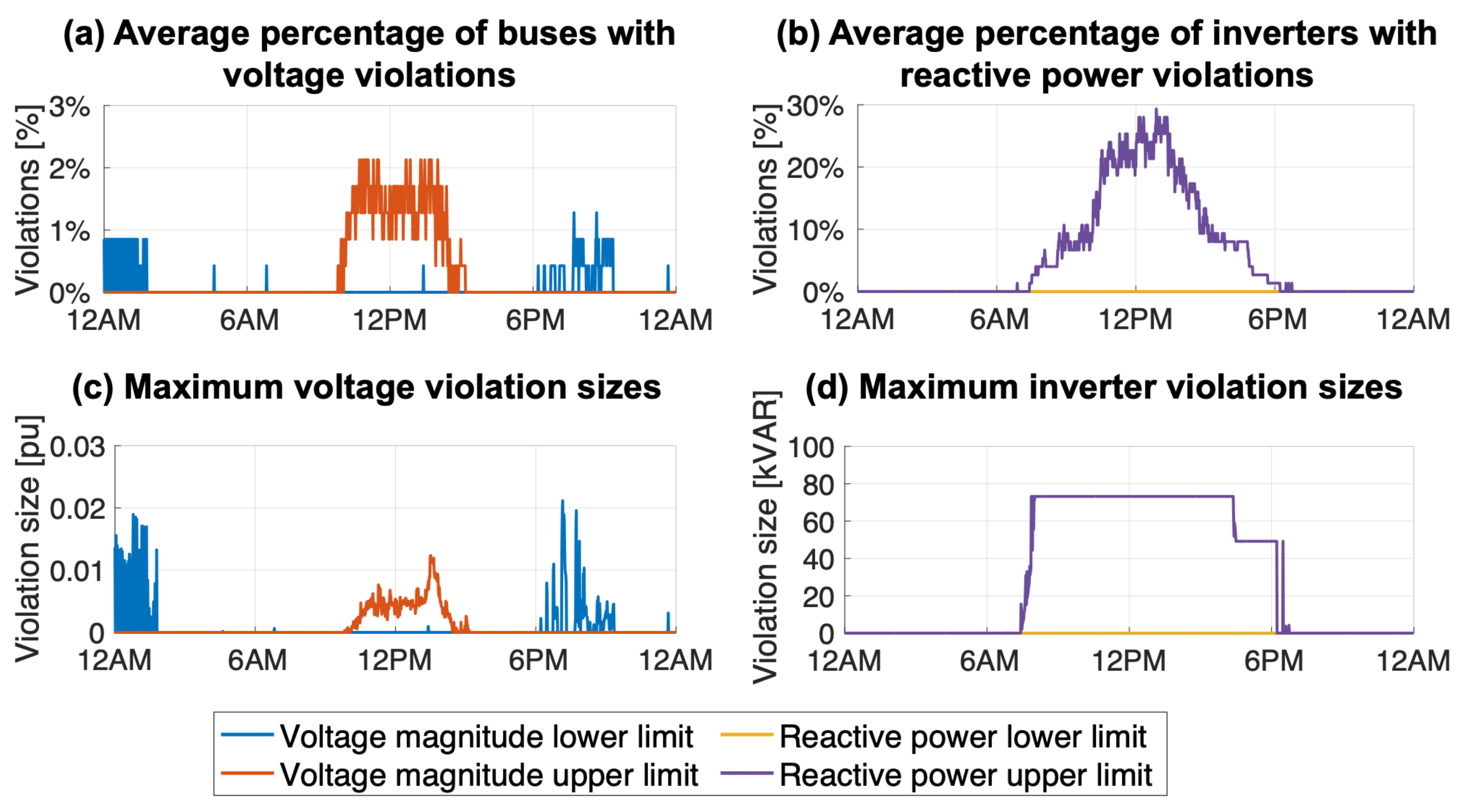}
	\vspace*{-8mm}
	\caption{Time-series plots for out-of-sample voltage magnitude and inverter reactive power limits violations for quantile-based method for single replication using~$M=2880$ random samples. The top plots show average number of violations throughout the entire day. The bottom plots illustrate the size of maximum violation throughout the entire day.}
	\label{fig:QM_time_series_trial5}
\end{figure}

\begin{figure}[!t]
	\centering	        	
	\includegraphics[width=0.49\textwidth]{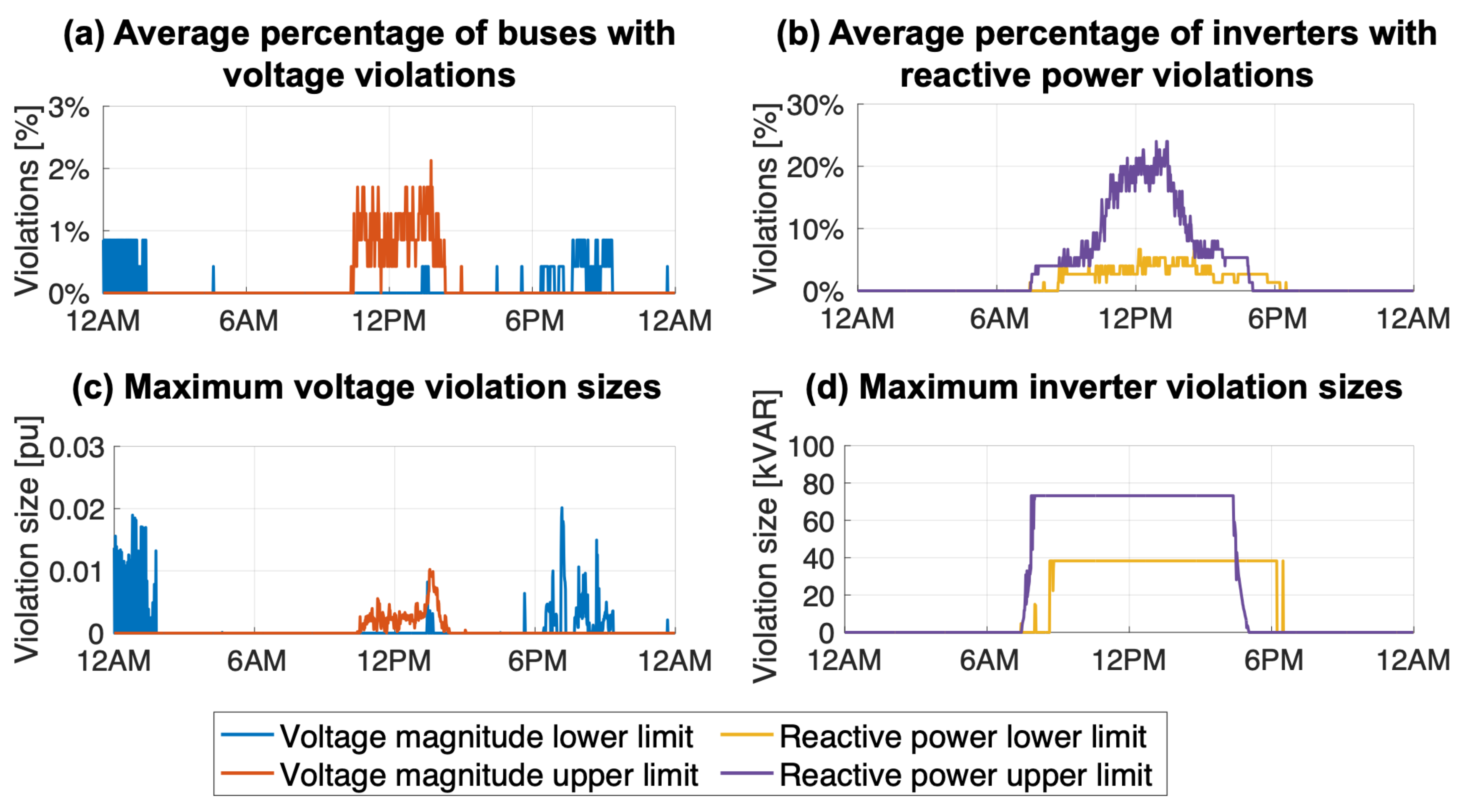}
	\vspace*{-8mm}
	\caption{Time-series plots for out-of-sample voltage magnitude and inverter reactive power limits violations for tuning-based method for single replication using~$M=2880$ random samples. The top plots show average number of violations throughout the entire day. The bottom plots illustrate the size of maximum violation throughout the entire day.}
	\label{fig:TM_time_series_trial5}
\end{figure}

We observed above 
that there are considerable voltage magnitude and inverter limit violations in the out-of-sample results. 
Since the out-of-sample evaluation is performed using full day data, we can analyze how the violations vary across the day. Fig.~\ref{fig:QM_time_series_trial5} shows violations over time for the quantile-based solution. The top plots show the percentage of single-phase nodes experiencing voltage violations (left) or inverters experiencing apparent power violations (right) at each time step of the day, averaged over all five of the out-of-sample evaluation days. The bottom plots show the magnitude of the worst case violation across all nodes or inverters and all evaluation days at each time step. In Fig.~\ref{fig:TM_time_series_trial5}, we plot the analogous results for the solution obtained with the tuning-based method.

In Fig. ~\ref{fig:QM_time_series_trial5}(a) and ~\ref{fig:TM_time_series_trial5}(a), the upper voltage violations (orange lines) typically occur around mid-day when the PV generation variability is high, whereas the lower voltage limit violations (blue lines) occur in the evening (between 6pm and 9pm) and at night (between midnight and 2am) when the load variability is high. However, we can see in Figures~\ref{fig:QM_time_series_trial5}(c) and ~\ref{fig:TM_time_series_trial5}(c) that the magnitude of either voltage magnitude violations are relatively small, with a maximum of around $0.02$ p.u. 

With respect to the inverter limits shown in Fig.~\ref{fig:QM_time_series_trial5}(b) and ~\ref{fig:TM_time_series_trial5}(b), both methods have inverter upper limit violations (purple lines) occurring across the entirety of the day-time when the PV generation is non-zero. In contrast to the voltage violation results, the percentage of nodes experiencing violations is relatively high, peaking at around $30\%$ around 1pm for the quantile-based method and $24\%$ for the tuning-based method. A noticeable difference between the two methods is that tuning-based method yields a set point with inverter lower limit violations throughout the day-time, whereas no such violations occur with quantile-based method results. The maximum magnitude of inverter upper limit violation is high for both methods in Fig.~\ref{fig:QM_time_series_trial5}(d) and ~\ref{fig:TM_time_series_trial5}(d), at $73.2$ kVAR throughout the almost entirety of the day-time period. The maximum magnitude of inverter lower limit violation for the tuning-based method is also high at $38.3$ kVAR.

\section{Case Study III: Considering the Impact of Inverter Capping}
\label{sec:case-study-2}

A strategy to mitigate inverter violations is capping the inverter reactive power injections to comply with the IEEE 1547-2018 standard~\cite{photovoltaics2018ieee}. In this section, we cap the inverter reactive power injections to their corresponding limits during the solution evaluation step described in Section \ref{sec:evaluation}. We use the same set of $M = 2880$ randomly sampled data points as in Section \ref{sec:single-rep-results} and investigate the impact of inverter capping on the voltage violations and total VUF. For all results in this section, we show results only for the quantile-based method since the results of the tuning-based method are very similar.

\subsubsection{Comparison of empirical distributions}
\begin{figure}[!t]
	\centering	        	
	\includegraphics[width=0.49\textwidth]{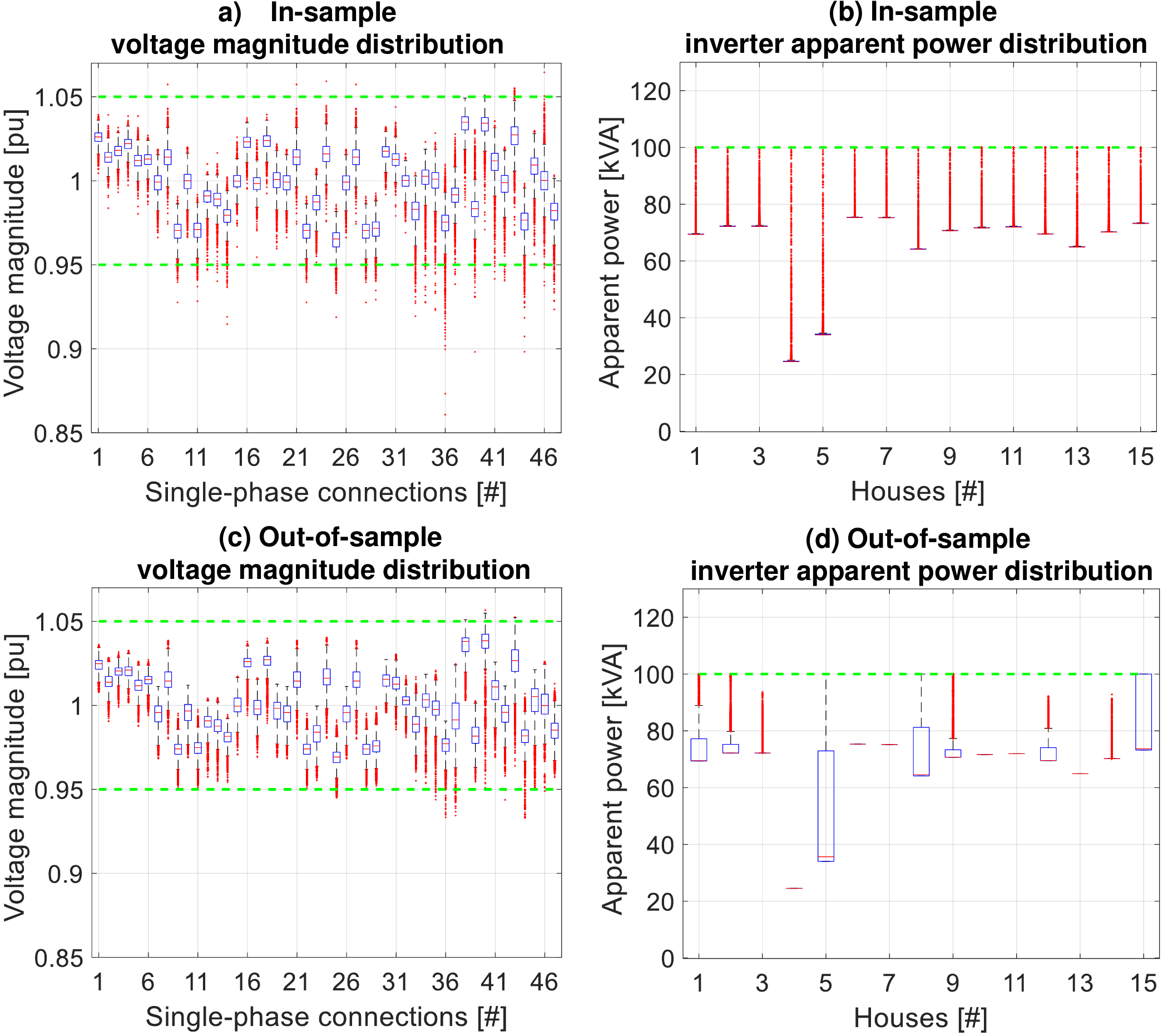}
	\caption{Box-whisker plots for voltage magnitude and inverter apparent power distribution calculated using Monte Carlo simulations with capping for the quantile-based method for single replication using~$M=2880$ random samples with inverter capping. The green dashed lines represent the lower and upper voltage magnitude limits and the maximum inverter apparent power limits.}
	\label{fig:QM_distr_trial5_capping}
\end{figure}

We first examine the impact of capping the inverter reactive power injections on the voltage and inverter violations. Shown in Fig.~\ref{fig:QM_distr_trial5_capping} are the voltage magnitude (left) and inverter apparent power (right) distributions for both in-sample (top) and out-of-sample (bottom) data. As expected, we do not observe any inverter limit violations in Figures~\ref{fig:QM_distr_trial5_capping}(b) and~\ref{fig:QM_distr_trial5_capping}(d) due to the capping. By comparing the voltage distribution plots in Fig.~\ref{fig:QM_distr_trial5_capping}(c) to the results obtained in Figs.~\ref{fig:QM_distr_trial5}(c) without capping, we see that there are less violations of the upper voltage magnitude limit (at nodes 38 and 40) and smaller violation sizes for the lower voltage magnitude limit (nodes 36 and 44) for the out-of-sample data. The decrease in the number of overvoltages and undervoltage violation size is likely due to to the reduced availability of reactive power injections because of capping.

Recall that while the reactive power injections are capped in the solution evaluation step, the reactive power limits in CCR-OPF are tightened using the~$\epsilon_q$ and~$1-\epsilon_q$ quantiles of the empirical distribution. So, we next investigate how changing~$\epsilon_q$ impacts the constraint tightenings, resulting VUF, and out-of-sample voltage magnitude violations.

\subsubsection{Impact of higher violation probability}
\begin{figure}[!t]
	\centering	        	
	\includegraphics[width=0.32\textwidth]{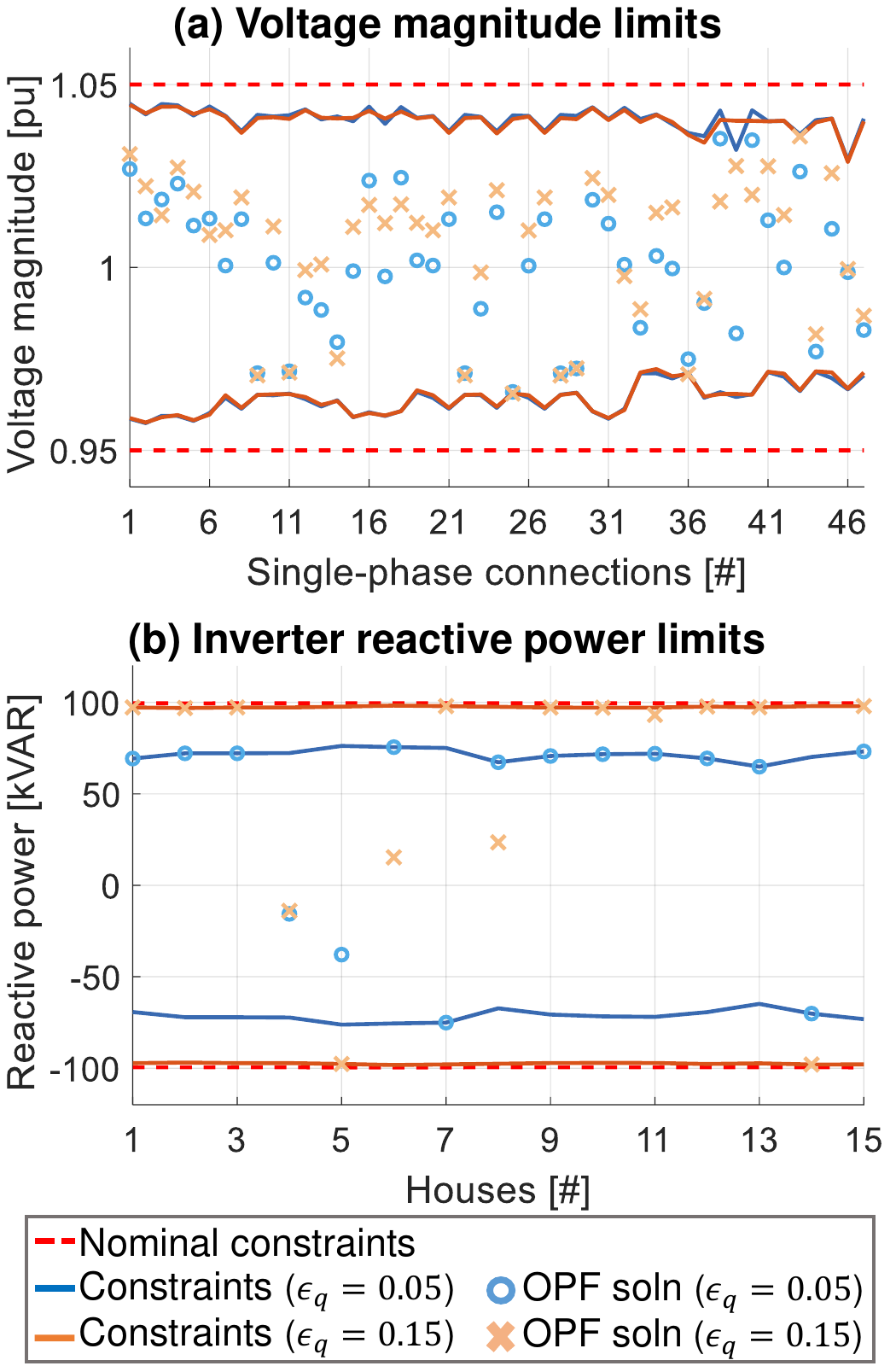}
	\caption{Comparison of CCR-OPF solution and constraint tightenings resulting from the quantile-based with capping for single replication using~$M=2880$ random samples with different~$\epsilon_q$.}
	\label{fig:constr_trial5_capping}
\end{figure}

We assess the impact of using a higher desired violation for the reactive power constraints. Here, we keep $\epsilon_v = 0.05$ while using $\epsilon_q=0.15$ in the quantile-based algorithm with reactive power capping. Fig.~\ref{fig:constr_trial5_capping} compares the constraint tightenings and operating points for~$\epsilon_q=0.05$ (blue lines) and~$\epsilon_q=0.15$ (orange lines). The voltage magnitude tightenings in Fig.~\ref{fig:constr_trial5_capping}(a) remain almost identical since we use~$\epsilon_v=0.05$ for both cases. However, by choosing a higher violation probability~$\epsilon_q=0.15$, we obtain a wider range for the reactive power limits in Fig.~\ref{fig:constr_trial5_capping}(b), which are very close to the nominal constraints (red dashed lines).

\begin{table}[!b]
\scriptsize
\begin{center}
\caption{In-sample results for quantile-based method with  capping for single replication with~$2880$ random samples.} 
\begin{adjustbox}{width=\columnwidth}
\begin{tabular}{c|c|c|c|c|c}
\toprule
{Case} & $\hat{E}_{\underline{v}}^{\max}$ & $\hat{E}_{\overline{v}}^{\max}$ & $\hat{E}_{\underline{q}}^{\max}$ & $\hat{E}_{\overline{q}}^{\max}$ & {VUF (\%)} \\
\midrule\midrule
No capping  & 0.05 & 0.005 & 0.05  & 0.05 & 6.32  \\
Capping, $\epsilon_q$=\,0.05 & 0.05  & 0.007 & 0.0 & 0.0 & 6.61  \\
Capping, $\epsilon_q$=\,0.15  & 0.05 & 0.05 & 0.0 & 0.0 & 5.85  \\
\bottomrule  
\end{tabular}
\end{adjustbox}
\label{tab:VUF_in-sample_capping}
\end{center}
\end{table}

\begin{table}[!b]
\scriptsize
\begin{center}
\caption{Out-of-sample results for quantile-based method with  capping for single replication with~$2880$ random samples.} 
\begin{adjustbox}{width=\columnwidth}
\begin{tabular}{c|c|c|c|c|c}
\toprule
{Case} & $\hat{E}_{\underline{v}}^{\max}$ & $\hat{E}_{\overline{v}}^{\max}$ & $\hat{E}_{\underline{q}}^{\max}$ & $\hat{E}_{\overline{q}}^{\max}$ & {VUF (\%)} \\
\midrule\midrule
No capping  & 0.019 & 0.126 & 0.0  & 0.36 & 6.97  \\
Capping, $\epsilon_q$=\,0.05 & 0.019  & 0.016 & 0.0 & 0.0 & 6.81 \\
Capping, $\epsilon_q$=\,0.15  & 0.018 & 0.029 & 0.0 & 0.0 & 6.21  \\
\bottomrule  
\end{tabular}
\end{adjustbox}
\label{tab:VUF_out-sample_capping}
\end{center}
\end{table}

Tables~\ref{tab:VUF_in-sample_capping} and~~\ref{tab:VUF_out-sample_capping} summarize the in- and out-of-sample results for a single replication of the quantile-based method using $\epsilon_q = 0.05$ without capping (from Section~\ref{sec:single-rep-results}), $\epsilon_q = 0.05$ with capping, and $\epsilon_q = 0.15$ with capping. All results use the same sample set ($M = 2880$ randomly drawn samples).
We compare the worst-case violation probabilities for all constraint types and the VUF. In Table~\ref{tab:VUF_in-sample_capping}, we observe that none of the in-sample violation probabilities in any of the three cases exceed $0.05$. Furthermore, when inverter capping is used, the violation probabilities for the inverter limits~$\hat{E}_{\underline{q}}^{\max},\hat{E}_{\overline{q}}^{\max}$ are zero. By capping with~$\epsilon_q=0.05$, there is a slight increase in the VUF compared to the VUF for case without capping. This increase is likely due to the limited reactive power support resulting from capping. By setting~$\epsilon_q=0.15$, we increase the feasible space for reactive power injections, leading to a better solution with the lowest VUF value among all three cases.

A similar trend can be observed in the out-of-sample results shown in Table~\ref{tab:VUF_out-sample_capping}. For the case without capping, we see that the out-of-sample violation probabilities for the inverter upper limits~$\hat{E}_{\overline{q}}^{\max}$ and voltage upper limits~$\hat{E}_{\overline{v}}^{\max}$ are higher than our desired violation probability of~$0.05$. By capping with~$\epsilon_q=0.05$, we are able to achieve violation probabilities that are considerably below~$0.05$. By setting~$\epsilon_q=0.15$, we obtain the lowest VUF value at the cost of increasing the voltage upper limit violations.

%% file: Sections/5_Conclusion.tex
\section{Conclusion}
\label{sec:conclusion}
In this paper, we develop iterative, data-driven algorithms for solving the chance-constrained AC OPF for unbalanced distribution grids. We reformulate the chance constraints into deterministic constraints consisting of the nominal constraints tightened with uncertainty margins. The optimal constraint tightenings are calculated using an iterative approach that alternates between solving a deterministic OPF with fixed tightenings and using sample-based evaluations to update the tightening terms. We propose two methods to perform the iterative updates: directly using the results of a Monte Carlo simulation (quantile-based method) or tuning using a safety parameter (tuning-based method).

Both methods were tested by running numerical simulations on the IEEE 13-bus test feeder using real residential load and PV data. Our case studies demonstrate that both methods perform in a similar manner and are able to enforce the chance constraints in the in-sample evaluation. The out-of-sample results were considerable improved by capping the DER power outputs. Furthermore, simulation results indicate that using randomly chosen samples across multiple days is the advantageous sampling procedure for both methods, as opposed to using representative full day sample sets.

For future work, we will focus on investigating the impact of using linear approximations of the power flow equations in order to solve the chance-constrained problem for large, realistic distribution feeders. We further plan to explore effective ways to identify a multi-dimensional tuning parameter for the tuning-based method so that we can separately tune the individual chance constraints, and will apply results from~\cite{hou2021data} to ensure that solutions obtained with the tuning-based methods provide rigorous feasibility guarantees.